\documentclass[onecolumn,a4paper,unpublished,11pt]{quantumarticle}
\pdfoutput=1

\usepackage[utf8]{inputenc}
\usepackage{graphicx}
\usepackage{hyperref}
\usepackage{amsmath}
\usepackage{amssymb}
\usepackage{xcolor}
\usepackage{amsthm}
\usepackage{subcaption}
\usepackage{cleveref}
\usepackage{tikz}
\usepackage{quantikz}
\usepackage[numbers,sort&compress]{natbib}

\newcommand{\kb}[2]{\ket{#1} \bra{#2}}

\newtheorem{theorem}{Theorem}

\title{Tangling schedules eases hardware connectivity requirements for quantum error correction}
\author{Gy\"orgy P. Geh\'er}
\email{gehergyuri@gmail.com, george.geher@riverlane.com}
\affiliation{Riverlane, St.~Andrew's House, 59 St.~Andrew's Street, Cambridge CB2 3BZ, United Kingdom}
\author{Ophelia Crawford}
\email{ophelia.crawford@riverlane.com}
\affiliation{Riverlane, St.~Andrew's House, 59 St.~Andrew's Street, Cambridge CB2 3BZ, United Kingdom}
\author{Earl T. Campbell}
\email{earl.campbell@riverlane.com}
\affiliation{Riverlane, St.~Andrew's House, 59 St.~Andrew's Street, Cambridge CB2 3BZ, United Kingdom}
\affiliation{Dept.~of Physics and Astronomy, University of Sheffield, Sheffield S3 7RH, United Kingdom}

\begin{document}

\maketitle

\begin{abstract}
   Error corrected quantum computers have the potential to change the way we solve computational problems. Quantum error correction involves repeated rounds of carefully scheduled gates to measure the stabilisers of a code. A set of scheduling rules are typically imposed on the order of gates to ensure the circuit can be rearranged into an equivalent circuit that can be easily seen to measure the stabilisers. In this work, we ask what would happen if we break these rules and instead use circuit schedules that we describe as tangled. We find that tangling schedules generates long-range entanglement not accessible using nearest-neighbour two-qubit gates. Our tangled schedules method provide a new tool for building quantum error correction circuits and we explore applications to design new architectures for fault-tolerant quantum computers. Notably, we show that, for the widely used Pauli-based model of computation (achieved by lattice surgery), this access to longer-range entanglement can reduce the device connectivity requirements, without compromising on circuit depth.
\end{abstract}


\section{Introduction}\label{sec:intro}
\begin{figure}[t]
    \centering
    \begin{subfigure}{0.45\textwidth}
        \centering
        \includegraphics[width=0.5\textwidth]{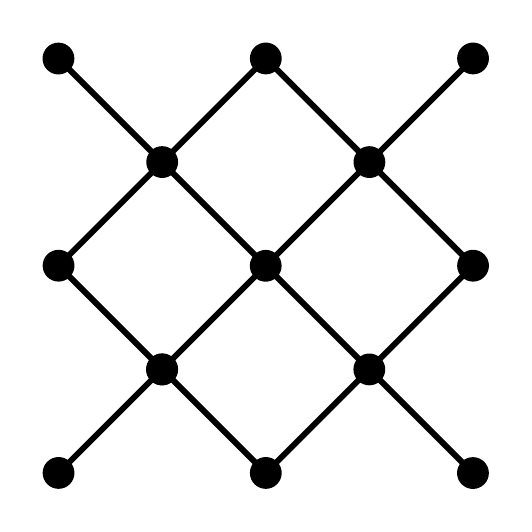}
        \caption{Rotated planar}
        \label{fig:rotplan}
    \end{subfigure}
    \begin{subfigure}{0.45\textwidth}
    \centering
        \includegraphics[width=0.5\textwidth]{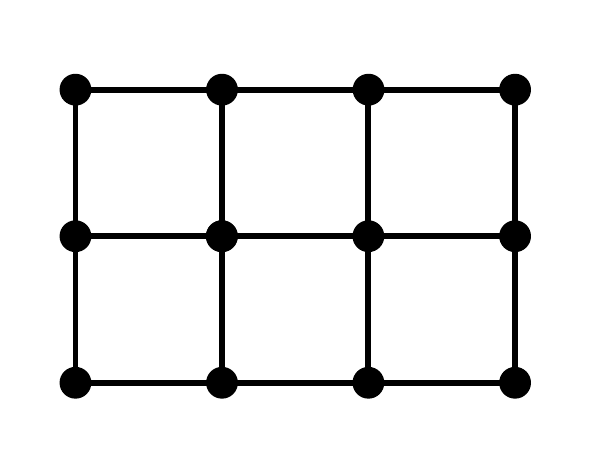}
        \caption{Unrotated planar}
    \end{subfigure}
    \caption{Hardware layouts with square-grid connectivity that accommodate (a) the rotated and (b) the unrotated planar codes, at least for the purpose of quantum memory.}\label{fig:planar_qpu}
\end{figure}

Building a quantum computer is a difficult task and physical qubits unfortunately tend to be noisy. As a result, running quantum algorithms without trying to correct errors introduced during physical execution gives very unreliable results. Quantum error correction (QEC) offers a solution by using several lower-quality physical qubits to encode a higher-fidelity logical qubit (see e.g. \cite{campbell2017roads}), so that algorithm outputs are more reliable. One approach to QEC uses stabiliser codes \cite{Gottesman-thesis}, where errors are detected by repeatedly measuring a set of stabilisers. One of the most popular types of stabiliser codes are the surface codes \cite{Kitaev-20032,Dennis-Kitaev-Landahl,Fow, RaussendorfHarrington}. They have relatively high threshold, and their stabilisers for the purpose of quantum memory can be measured easily on a square-grid layout hardware with only nearest-neighbour two-qubit gates, such as Google's Sycamore \cite{Google-Sycamore} (see also \Cref{fig:planar_qpu}).

However, to go beyond quantum memory and execute fault-tolerant quantum computation (FTQC), we need to perform logical gates on our logical qubits. Almost all modern resource estimates use the so-called Pauli-based computational (PBC) model \cite{BravyiPBC} for FTQC to estimate the space-time cost of error corrected algorithms \cite{PhysRevResearch.3.033055, PhysRevResearch.5.013200, PRXQuantum.2.030305,PhysRevX.8.041015}. The PBC model consists of measuring a series of multi-logical-qubit Pauli operators, which can be achieved on a $2$D planar architecture with the planar surface code via \emph{lattice surgery} \cite{Horsman_2012, LitOpp,GoSc,ChamCamtwistbased,ChamCamtwistfree,FowlerGidneyLS}. In particular, most resource estimates use Litinski's proposal from ``A Game of Surface Codes'' \cite{GoSc}; however, this ignores the microscopic details of hardware constraints.

Indeed, most types of quantum processing units (QPU), e.g. superconducting, typically have fixed qubit layout and connectivity. Here, connectivity means which pairs of qubits can be acted on with native two-qubit gates. As connectivity increases, so too does crosstalk noise and related engineering challenges. Furthermore, a uniform connectivity QPU is desirable so that different code sizes and algorithms can be executed on the same device. As such, uniform, low-degree QPUs are the natural choice. We already see these constraints in the design of current superconducting QPUs. For instance, IBM's Eagle and Rigetti's Aspen both have degree-three connectivity for all qubits in the bulk \cite{IBM-Eagle, Rigetti-Aspen, Rigetti-tunable-coupler}. Google's Sycamore \cite{Google-Sycamore} has the square-grid connectivity (\Cref{fig:planar_qpu}), on which we can naturally place planar code patches such that all stabilisers are local (i.e. one auxiliary qubit can be allocated for each that is connected to its data qubits). However, PBC requires the measurement of some irregularly-shaped, long-range stabilisers, typically so-called \emph{elongated rectangles} and \emph{twist defects}. A naive approach to measure these irregular stabilisers would be to swap qubits as needed, but this would increase circuit depth and thus would be very detrimental to logical fidelity and thresholds. The literature lacks a general technique that would achieve the measurement of such long-range stabilisers without compromising on the mentioned disadvantages.

In this paper, we propose \emph{tangled syndrome extraction} that enables measurement of long-range and/or high-weight stabilisers on a restricted connectivity device. Our method considers the target stabiliser as a product of smaller-weight \emph{component operators} that are local with respect to the QPU. We achieve this by starting with the naturally-arising syndrome extraction circuits that measure the component operators, \emph{tangling} these circuits, and changing the measurement basis for some of the auxiliary qubits in the last layer. The result is a syndrome extraction circuit that measures the product of the component operators. 

As a main application of our tangled syndrome extraction method, we show a solution to the above problem of \emph{measuring the elongated rectangles and twist defects under square-grid connectivity}. As a consequence, we present two ways to perform FTQC under degree-four connectivity, one with the unrotated planar code and another with the rotated planar code. We note that our tangled syndrome method has a range of applications beyond planar codes; however, since planar codes are a strong contender for FTQC, with the aforementioned gap in the literature, we shall mainly focus on them throughout the paper.

The paper is organised as follows. In the next section, we explain the standard scheduling rules that are usually applied to construct a circuit that measures a set of stabilisers simultaneously and independently. Then in \Cref{sec:our_method} we show in detail how breaking these rules for a pair of component operators, by tangling their syndrome extraction circuits, results in a circuit that measures their product. We then state our general theorem for an arbitrary number of component operators, which we prove in \Cref{app:proof_of_main_thm}. In \Cref{sec:elong-twistd}, we show how to use tangled syndrome extraction for measuring elongated-rectangle and twist-defect stabilisers, which involves the use of so-called \emph{accessory qubits} that are initialised in a $Y$-eigenstate. To demonstrate our tangled syndrome technique's performance, we present numerical results in \Cref{sec:numerics} that compare both the quantum memory and stability experiments \cite{Gidney2022stability} for the default planar code patch, that has only regular stabilisers, and a tangled version of the patch that contains some elongated rectangular stabilisers. We then show how to perform a general lattice surgery operation (i.e. twist-based lattice surgery) with the unrotated planar code on the square-grid connectivity QPU in \Cref{sec:full_lattice_surgery}, and discuss the case of the rotated planar code in \Cref{app:rotated_lattice_surgery}. In \Cref{sec:conclusion}, we summarise our results and discuss potential next steps. \Cref{app:a} contains further supplementary material and a discussion on existing alternative methods to execute FTQC without PBC on the square-grid connectivity QPU as in  \Cref{fig:planar_qpu}.

\begin{figure}[t]
    \centering
    \includegraphics{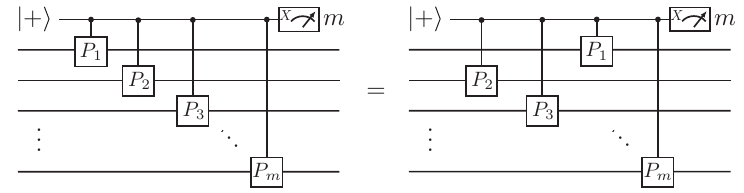}
     \caption{Two auxiliary circuits for measuring the stabiliser $P_1P_2\dots P_j$ using an auxiliary qubit prepared in the $|+\rangle$ state. The circuits differ only in their schedules and, assuming no other circuits are executed simultaneously, the two are equivalent. The measurements are in the X basis.}\label{fig:aux_circ}
\end{figure}

\section{Background}

This section explains the scheduling rules that are usually used to construct a circuit that measures a set of stabilisers simultaneously and independently. Our tangled syndrome extraction technique, explained in the next section, builds on these rules.

First, we recall some simple circuits for syndrome extraction. Consider a general stabiliser $g = P_1P_2\dots P_m$, where each of $P_1,P_2,\dots, P_m$ is a single-qubit Pauli operator, with no two of them acting on the same qubit.  We can always measure $g$ using the following circuit: prepare an auxiliary qubit (labelled $1$ below) in the $|+\rangle$ state; apply controlled-$P_{1}$; apply controlled-$P_{2}$; $\ldots$ apply controlled-$P_{m}$, each controlled on the auxiliary qubit; and finally, measure the auxiliary qubit (qubit $1$) in the $X$ basis. This measurement outcome corresponds to the measurement of $g$.  

Furthermore, we are free to shuffle the order of the controlled-$P_{j}$ gates and we may also insert identity gates. To specify this ordering, we define a schedule (one-to-one) map $f:[1,\ldots,m] \rightarrow [1,\ldots,t]$ where $t$ is the depth of the circuit. That is, in time step $i$, we should implement controlled-$P_{f^{-1}(i)}$ if $f^{-1}(i)$ exists. If the inverse does not exist, we write $f^{-1}(i)=\emptyset$ where $\emptyset$ denotes the empty set, and the gate $P_{\emptyset}$ is the identity $I$. Therefore, the whole protocol to measure $g$ is
\begin{itemize}
    \item step $0$: prepare qubit $1$ in the $|+\rangle$ state, 
    \item step $1$: apply controlled-$P_{f^{-1}(1)}$,
    \item \dots, 
    \item step $t$: apply controlled-$P_{f^{-1}(t)}$,
    \item step $t+1$: measure out qubit $1$ in the $X$ basis, the outcome corresponding to the measurement of $g$.
\end{itemize}
We call this the \emph{auxiliary syndrome extraction circuit of $g$} (see also \Cref{fig:aux_circ}). Once we have the auxiliary circuit, we can compile it to the native gate-set of the physical device (e.g. using only $CZ$ as the native two-qubit gate). The auxiliary syndrome extraction circuit is a standard tool, often used to measure the stabilisers of the surface code. We note that, for weight-$w$ stabilisers, a single qubit error on the auxiliary qubit can propagate to $w/2$ data qubits (up to stabiliser equivalence) and so, for high-weight stabilisers, it is often modified to use more auxiliary qubits (e.g. flag qubits) for syndrome extraction.

\begin{figure}[h]
    \centering
    \begin{subfigure}{0.45\textwidth}
        \centering
        \includegraphics[width=0.45\textwidth]{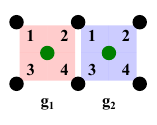}
        \caption{Scheduling that measures the stabilisers $g_1=XXXXII$ and $g_2=IIZZZZ$.}
        \label{fig:untng_plaqs1}
    \end{subfigure}
    \hspace{0.05\textwidth}
    \begin{subfigure}{0.45\textwidth}
        \centering
        \includegraphics[width=0.3\textwidth]{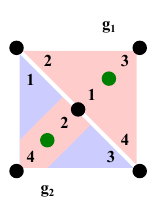}
        \caption{Scheduling that measures the stabilisers $g_1=XXXXI$ and $g_2=IZXZX$.}
        \label{fig:untng_plaqs2}
    \end{subfigure}
    \caption{Examples of schedules for pairs of stabilisers that satisfy condition (b). The colouring of qubits and plaquettes in this figure is used throughout the paper. Stabiliser $X$ and $Z$ Pauli terms are coloured red and blue, respectively. A continuous coloured area indicates a single stabiliser, with the black circles on the edge representing data qubits and the green circle within the shape the auxiliary qubit used for measurement of the stabiliser. Each number between a pair of data and auxiliary qubits indicates the layer in which the corresponding entangling gate between the two qubits is applied during syndrome extraction. In both examples, there are two joint qubits where the Pauli terms differ; specifically, for $g_1$, these are $X$s and for $g_2$ they are $Z$s. We therefore either need to apply the two $CX$ gates for $g_1$ before we apply the $CZ$ gates for $g_2$, or the other way around. Otherwise, we would not measure the two desired stabilisers.}\label{fig:2_plaqs_default}
\end{figure}

Next, consider a set of stabilisers $\{g_j\}_{j=1}^m$. The commutation relations ensure that, in theory, they can be measured simultaneously. This is indeed the case; however, there are additional restrictions on the schedules. Let us consider a schedule $f_j$ for each stabiliser $g_j$, defining the auxiliary syndrome extraction circuits $\{\mathcal{C}_j\}_{j=1}^m$. We assume they have the same depth, i.e. number of layers including those with identity gates, and that we use different auxiliary qubits for different stabilisers. If we combine these circuits, so that they all occur simultaneously, the resulting circuit, which we denote by $\mathcal{C}$, then has to satisfy the following conditions to achieve the desired behaviour:
\begin{itemize}
    \item[(a)] No qubit is involved in more than one gate at a time. 
    \item[(b)] For every pair of distinct circuits $j\neq k $, the simultaneous combination of $\mathcal{C}_j$ and $\mathcal{C}_k$ is equivalent to the serial execution $\mathcal{C}_j$ followed by $\mathcal{C}_k$. 
\end{itemize}
These conditions are equivalent to the following more formal statements
\begin{itemize}
    \item[(a')] For every time step index $i$ and every pair of distinct schedules $j \neq k$, we have that $f_j^{-1}(i) \cap f_k^{-1}(i) = \emptyset$.
    \item[(b')] Let $g_j=P^{j}_1 P^{j}_2\ldots$ and $g_k=P^k_1 P^k_2\ldots$ and consider the set $\mathcal{K}_{j,k}$ of anti-commuting pairs $\{ P^{j}_{\alpha} , P^{k}_{\beta} \}=0$.  Then we require that $f_j(\alpha) < f_k(\beta)$ for an even number of these anti-commuting pairs.
\end{itemize}
The equivalence of (b) and (b')  can be seen, for instance, by using the interchanging identity depicted in \Cref{fig:CX_CZ_interchange}. Conditions (a) and (b) are illustrated e.g. in \cite[Fig. 15 b-c]{LitOpp}, and we also depict two examples in \Cref{fig:2_plaqs_default}.

\section{Our method}\label{sec:our_method}

Now, we are ready to introduce our tangled schedules technique. We say that the circuits $\mathcal{C}_j$ and $\mathcal{C}_k$ are \emph{tangled} if condition (a) is satisfied but condition (b) is not. We will show that if $\{\mathcal{C}_j\}_{j=1}^m$ contains tangled pairs that satisfy certain conditions (detailed below), and we change the bases of measurements in their combined circuit $\mathcal{C}$, then the resulting circuit $\tilde{\mathcal{C}}$ measures the product $h=g_1\cdots g_m$. Since, in this case, the operators $g_j$ are no longer stabilisers themselves, we emphasise this by calling them \emph{component operators} instead. We start by explaining how this protocol works for a product of two component operators. Then, we state the general method involving an arbitrary number of component operators.

\begin{figure}[h]
    \centering
    \begin{subfigure}{0.45\textwidth}
    \centering
        \includegraphics[width=0.45\textwidth]{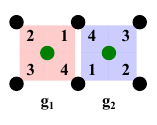}
        \caption{Tangled scheduling that measures the stabiliser $h = g_{1}g_{2} = -XXYYZZ$.}
        \label{subfig:untng_plaqs1}
    \end{subfigure}
    \hspace{0.05\textwidth}
    \begin{subfigure}{0.45\textwidth}
        \centering
        \includegraphics[width=0.3\textwidth]{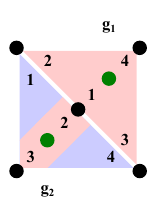}
        \caption{Tangled scheduling that measures the stabiliser $h = g_1 g_2 = -XYIYX$.}
        \label{subfig:untng_plaqs2}
    \end{subfigure}
    \caption{Examples of schedules that violate condition (b) and hence the combined circuits do not measure $g_1$ and $g_2$. However, we can exploit this to measure the higher-weight product, $h = g_{1}g_{2}$. For an explanation of the different parts of the diagrams, see~\Cref{fig:2_plaqs_default}; however, here, each continuous coloured area indicates a single component operator.}
    \label{fig:2_plaqs_tangled}
\end{figure}

Let us assume that $g_1$ and $g_2$ are two component operators that commute, and call their auxiliary qubits $1$ and $2$, respectively. Without loss of generality (i.e. by applying local Clifford equivalence), we may assume that the Pauli terms of $g_1$ are all of $X$ type, and those Pauli terms of $g_2$ that anticommute with any Pauli term of $g_1$ are of $Z$ type. Note that, as the components commute, the number of qubits where the Pauli terms anticommute is even. Let us assume we scheduled $g_1$, $g_2$ in a tangled way, i.e. condition (a) is satisfied but condition (b) is not; see \Cref{fig:2_plaqs_tangled} for two examples. Now, we claim that if we combine the two auxiliary syndrome extractions $\mathcal{C}_1$, $\mathcal{C}_2$ and replace the $X$-basis measurements with $Y$-basis measurements, then the resulting circuit $\tilde{\mathcal{C}}$ measures the product $h=g_1g_2$, and the measurement outcome is the mod-2 sum of the two auxiliary qubit outcomes. Furthermore, if we apply the Clifford correction $\sqrt{g_1}g_1^{m_1}$, where $m_1$ is the outcome on auxiliary qubit $1$ and $\sqrt{g_1} = \frac{1}{\sqrt{2}}(I+ig_1)$, then on the data qubits we have the post-measurement state. Therefore $\tilde{\mathcal{C}}$ measures the stabiliser $h=g_1 g_2$. We will also see that, if we execute this circuit twice, the two Clifford corrections combine into a Pauli correction that can be tracked in software. These circuits are illustrated in \Cref{fig:2_plaqs_circ} and \Cref{fig:2_plaqs_circ_2_rounds}. 

\begin{figure}[h]
    \centering
    \begin{subfigure}[t]{0.39\textwidth}
        \includegraphics[width=\textwidth]{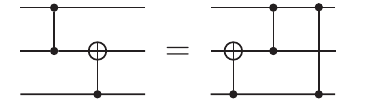} \caption{} \label{fig:CX_CZ_interchange}
    \end{subfigure}
    \hfill
    \begin{subfigure}[t]{0.39\textwidth}
        \includegraphics[width=\textwidth]{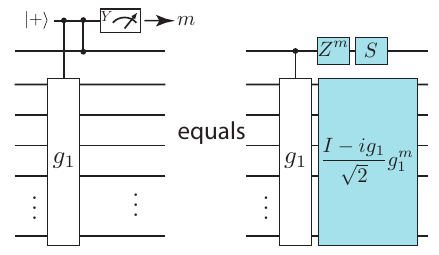} \caption{}  \label{Fig:Prune}
    \end{subfigure}
    \caption{Circuit identities used to analyze tangled schedules. (a) Interchanging rule for non-commuting entangling gates $CX$ and $CZ$. Similar rules hold for $CX, CY$ and $CY,CZ$ gate pairs, provided they share the target qubit. (b) The circuit \textit{pruning} identity for removing an entangled auxiliary qubit. In the text, we label the top qubit 1 and the next-to-top qubit 2. The blue boxes indicate Clifford corrections.  Note only the Pauli terms depend on the measurement outcome $m$, so that the (non-Pauli) Clifford term can be applied non-adaptively and the Pauli correction can also be handled non-adaptively by Pauli frame tracking. An application of this identity is \Cref{fig:2_plaqs_circ}. }
\end{figure}

\begin{figure}[h]
    \centering
    \includegraphics[width=0.8\textwidth]{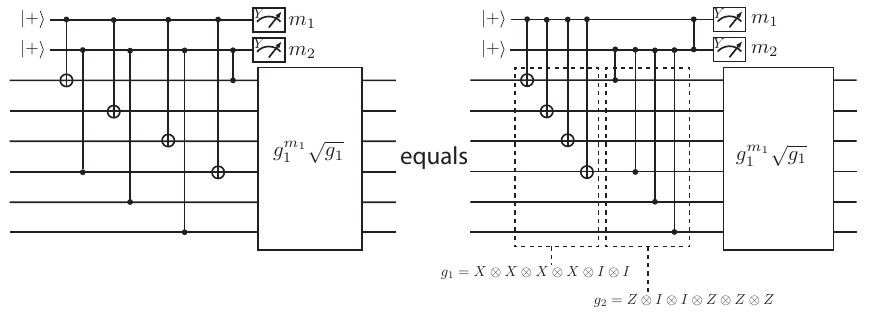}
    \caption{Circuit for measuring $g_1g_2= - XXYYZZ$ with the outcome given by $m_1 \oplus m_2$.  (left of equality): A circuit requiring low connectivity to measure the product $g_1 g_2$ by violating the (b) condition. This uses the schedule in \Cref{subfig:untng_plaqs1}. (right of equality): An equivalent circuit that shows direct entanglement of auxiliary qubits. In both cases, there is a Clifford correction to achieve the desired post-measurement state. After two rounds of the protocol, the Clifford correction becomes a Pauli correction as shown in \Cref{fig:2_plaqs_circ_2_rounds}.}\label{fig:2_plaqs_circ}
\end{figure}

Next, we prove that $\tilde{\mathcal{C}}$ indeed measures the product $g_1g_2$. Note the (multiplicative) commutator rules 
\begin{equation}\label{eq:commutator_rules}
[CX_{c_1,t},CY_{c_2,t}] = [CX_{c_1,t},CZ_{c_2,t}] = [CY_{c_1,t},CZ_{c_2,t}] = CZ_{c_1,c_2}   
\end{equation}
depicted in \Cref{fig:CX_CZ_interchange}. In other words, interchanging the order of two non-commuting controlled-Pauli gates has the effect of entangling the control qubits with a $CZ$ gate. \emph{This entangling of auxiliary qubits is the main feature of schedule tangling} that we will exploit, as shown in \Cref{fig:2_plaqs_circ}. 

Before proceeding, we prove the circuit pruning identity, shown in \Cref{Fig:Prune}. We define $C_i(g_j)$ to be a controlled $g_j$ gate with qubit $i$ as the control. Using $\ket{\phi}$ to describe the initial state of all qubits except the first auxillary qubit (qubit $1$), we have
\begin{align}
    \ket{\Psi} =  CZ_{1,2} \, C_1(g_1)\ket{+}_1 \ket{\phi} = \frac{1}{\sqrt{2}}(\ket{0}_1 + \ket{1}_1Z_2 g_1 )\ket{\phi}.
\end{align}
Performing a measurement in the $Y$ basis on qubit 1 with outcome $m$ leads to a projection by $\Pi_m = \ket{i_m}_{1}\bra{i_m}_{1}$ where $\ket{i_m}:=(\ket{0} + i (-1)^m\ket{1} )/\sqrt{2}$. The final state is therefore proportional to
\begin{align}
   \Pi_m \ket{\Psi} & \propto \ket{i_m}_{1}[\langle i_m \vert 0 \rangle_{1} + \langle i_m \vert 1  \rangle_{1} Z_2 g_1 ]\ket{\phi} \\ \nonumber
   & \propto \ket{i_m}_{1}[I - i (-1)^{m}Z_2 g_1 ]\ket{\phi}.
\end{align}
One can then verify that
\begin{align} 
     I - i (-1)^{m}Z_2 g_1 & = (\kb{0}{0}_2 + \kb{1}{1}_2) -  i (-1)^m (\kb{0}{0}_2 - \kb{1}{1}_2)g_1 \nonumber \\
     & = \kb{0}{0}_2 [I - i(-1)^m g_1] + \kb{1}{1}_2[I + i(-1)^m g_1] \nonumber \\
     & = \kb{0}{0}_2 [I - i(-1)^m g_1] + \kb{1}{1}_2 i(-1)^m g_1 [I - i(-1)^m g_1] \nonumber \\
     & = S_2 Z_2^m C_2(g_1) [I - i(-1)^m g_1] \nonumber \\
        & \propto S_2 Z_2^m C_2(g_1) (I - i g_1) g_1^m 
\end{align}
where $S=\kb{0}{0}+ i \kb{1}{1}$ is the standard phase gate. This proves the identity of \Cref{Fig:Prune}.  

We will now use the pruning identity to prove the efficacy of our tangled schedules technique in the case of a two-component stabiliser. Given two auxiliary qubits, 1 and 2, that have been entangled by a tangled schedule, with component operators $g_1$ and $g_2$, the state prior to measurement is
\begin{equation}
    \ket{\Psi} =  CZ_{1,2} \, C_1(g_1) C_2(g_2)\ket{+}_1 \ket{+}_2 \ket{\psi},
\end{equation}
where $\ket{\psi}$ denotes the initial state on the data qubits. Using the pruning identity (\Cref{Fig:Prune}) with qubit 1 as the first auxiliary qubit and obtaining measurement outcome $m_1$, we find that the state on the remaining qubits is
\begin{align}
\ket{\Psi'} & = \frac{1}{\sqrt{2}} S_2 Z_2^{m_1}C_2(g_1) C_2(g_2)\ket{+}_2 (I - i g_1) g_1^{m_1} \ket{\psi} \\
 & = \frac{1}{2} (\ket{0}_2 + i(-1)^{m_1}\ket{1}_2g_1g_2 )(I - i g_1) g_1^{m_1} \ket{\psi}.
\end{align}
Now, measuring out qubit $2$ in the $Y$ basis with outcome $m_2$ leads to a projection $\kb{i_{m_2}}{i_{m_2}}_2$. Therefore, we obtain the following state
\begin{align}
\ket{\Psi''} & \propto \kb{i_{m_2}}{i_{m_2}}_2\ket{\Psi'} \\ \nonumber
& \propto \ket{i_{m_2}}_2 (\langle i_{m_2}\ket{0}_2 + i(-1)^{m_1}\langle i_{m_2} \ket{1}_2 g_1g_2 )(I - i g_1) g_1^{m_1} \ket{\psi}  \\ \nonumber
& \propto \ket{i_{m_2}}_2  \left( \frac{I +(-1)^{m_1+m_2}g_1g_2}{2} \right)\left(\frac{I - i g_1}{\sqrt{2}} \right)g_1^{m_1}  \ket{\psi}. \nonumber
\end{align}
The first bracketed factor is a projector onto the $(-1)^{m_1+m_2}$ eigenspace of $g_1g_2$, thus measuring the desired operator. The second bracketed factor is a Clifford unitary $V$ such that (up to global phase) $V^2=g_1$, while the third factor is a Pauli unitary. It is straightforward to see that, in order to obtain the post-measurement state $\frac{1}{2}(I +(-1)^{m_1+m_2}g_1g_2 )\ket{\psi}$, we need to apply the Clifford correction $\frac{1}{\sqrt{2}}(I - i g_1)^\dagger g_1^{m_1}$ on the data qubits. Therefore, after two rounds, the full correction becomes Pauli. Indeed, a simple calculation shows that the full correction after the second round is $g_1^{m_1+n_1+1}$, where $n_1, n_2$ are the measurement outcomes in the second round; see also \Cref{fig:2_plaqs_circ_2_rounds}.

\begin{figure}
    \centering
        \includegraphics[width=0.7\textwidth]{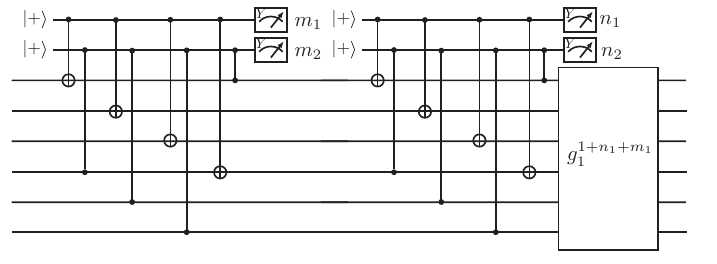}
    \caption{Circuit for two rounds of measuring $g_1g_2= - XXYYZZ$ with the outcomes given by $m_1 \oplus m_2$ and  $n_1 \oplus n_2$.  The Clifford correction from the one-round circuit in \Cref{fig:2_plaqs_circ} has become a Pauli correction $g_1^{1+m_1+n_1}$ that does not need to be physically implemented (removing the need for fast feedback) and can instead be accounted for by Pauli frame tracking.} \label{fig:2_plaqs_circ_2_rounds}
\end{figure}

Having proved the simplest example, we now describe our general protocol for measuring products of sets of pair-wise commuting component operators.

\begin{theorem}\label{thm:main}
Consider a set of pair-wise commuting Pauli product operators $\{g_j\}_{j=1}^m$ and a scheduling $\{f_j\}_{j=1}^m$ for each that defines their auxiliary syndrome extraction circuits $\{\mathcal{C}_j\}_{j=1}^m$. Denote by $\mathcal{C}$ the combined circuit. Compose an (undirected) graph $G=(V,E)$ where $V=[1, \dots, m]$ and $(j, k) \in E$ if and only if the schedules $f_j$ and $f_k$ are tangled. Suppose further that $G$ is a forest whose connected (tree) components are $\mathcal{T}_1, \dots, \mathcal{T}_\ell$. 
Then there exists a modification of $\mathcal{C}$ where
\begin{itemize}
    \item we modify the single qubit Pauli measurements on the auxiliary qubits, and
    \item we apply a Clifford correction on the data qubits,
\end{itemize}
such that the modified circuit $\tilde{\mathcal{C}}$ measures the products $h_r := \prod_{j\in\mathcal{T}_r} g_j$ for $r=1,\dots,\ell$ simultaneously and independently. Moreover, after two rounds of syndrome extraction $\tilde{\mathcal{C}}$, the accumulated Clifford corrections multiply into a Pauli correction that can be tracked in software.
\end{theorem}
We prove this general theorem in \Cref{app:proof_of_main_thm}. Here, we give a very brief proof sketch. For a simple two-vertex tree, we have already given a proof above. For a larger tree, we can iteratively identify leaves of the tree (vertices connected by only one edge) and remove them by using the pruning identity in \Cref{Fig:Prune}. We continue with this process until the remaining tree becomes trivial. It is also straightforward to see that each iterative correction is Clifford and commutes with all component operators, and therefore the combined correction after one round coming from all the trees is also Clifford. After an even number of rounds, the structure of each iterative correction shows that the combined two round correction is entirely Pauli, hence can be tracked in software.

In the next few sections, we give applications for the case when the forest $G$ contains only two-element trees apart from isolated vertices, which is especially interesting for applications in lattice surgery schemes. We shall leave applications of the more general case to future work.


\section{Measuring long-range stabilisers on a square-grid connectivity device}\label{sec:elong-twistd}

In the PBC model, we perform a series of multi-logical-qubit Pauli measurements while consuming a supply of magic states \cite{bravyi2005universal,GoSc,ChamCamtwistfree}. In the case of the planar code, a well-established method to do this is via lattice surgery \cite{Horsman_2012,LitOpp,GoSc,ChamCamtwistbased,ChamCamtwistfree,FowlerGidneyLS}. There are two types of lattice surgery we distinguish between: lattice surgery that merges logical patches only along their $X$ or $Z$ boundaries, and lattice surgery that may merge some involved patches along both of their boundaries. If the multi-logical-qubit Pauli operator does not involve any $Y$ Pauli terms but only $X$ and $Z$, then we can perform this measurement via one lattice surgery operation of the former type. In the case when the Pauli operator does involve at least one $Y$ Pauli term, we either need two lattice surgery operations of the former type (twist-free lattice surgery \cite{ChamCamtwistfree}), or one lattice surgery operation of the second type (twist-based lattice surgery \cite{ChamCamtwistbased}) to measure the Pauli product. Most stabilisers during either type of lattice surgery are local, i.e. can be measured with their auxiliary syndrome extraction circuits on the natural square-grid connectivity QPU (\Cref{fig:planar_qpu}).  However, in both cases, we typically need to measure a few irregularly-shaped long-range stabilisers too. A naive solution would be to swap qubits around to generate the required entanglement, but this would significantly deepen syndrome extraction circuits leading to lower thresholds and longer computations.   We are not aware of any prior proposal (with low depth) to measure these long-range stabilisers needed for fully-general lattice surgery, and instead prior art either proposes QPU modifications or ignores the problem. For instance, Ref. \cite{ChamCamtwistfree} modifies the QPU of \Cref{fig:planar_qpu} 
in an area by introducing so-called dislocations, Ref. \cite{ChamCamtwistbased} increases the connectivity from four to six, while Refs \cite{LitOpp,GoSc} do not address this issue.

Here, we show how to measure the irregularly-shaped long-range stabilisers on the natural square-grid connectivity QPU using tangled schedules. There are two types of long-range stabilisers we need to consider: \emph{elongated rectangles} and \emph{twist defects}. To motivate their roles, first let us consider a square-grid connectivity QPU on which we have placed two planar code patches. Suppose further that we can perform $XX$-type lattice surgery between them using only local stabilisers. In this case, it is straightforward to see that $ZZ$-lattice surgery is also possible (given enough qubits around that can be used for the merge stage); however, the mixed $XZ$- and $ZX$-type lattice surgeries are not possible without using long-range stabilisers. Indeed, this can be seen via a colouring argument which we explain in \Cref{app:background-lattice}. In such a case, we say that the two patches are aligned with respect to the background lattice; otherwise, i.e. when $XZ$ lattice surgery is possible using only local stabilisers but $XX$ and $ZZ$ lattice surgeries require long-range stabilisers, we say they are anti-aligned. 

\begin{figure}[h]
    \centering
    \begin{subfigure}{0.45\textwidth}
        \centering
        \includegraphics[width=0.5\textwidth]{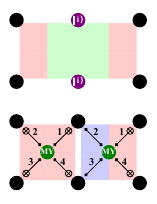}
        \caption{}
        \label{fig:elongated1}
    \end{subfigure}
    \hspace{0.05\textwidth}
    \begin{subfigure}{0.45\textwidth}
        \centering
        \includegraphics[width=0.5\textwidth]{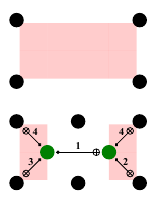}
        \caption{}
        \label{fig:elongated2}
    \end{subfigure}
    \caption{Measuring an elongated $XXXX$ stabiliser (a) using our method that requires only a degree-four connectivity device and (b) using the method of \cite{ChamCamtwistbased} that requires a degree-six connectivity device. With our method, we initialise the two (accessory) qubits in the middle in the $Y$ basis (in purple), then measure the product $XXXXII\cdot IIZZXX = -XXYYXX$ of the two plaquettes by using schedules that do not satisfy condition (b). The classically flipped outcome of this measurement is then the outcome corresponding to the elongated rectangle. Here and in the rest of the paper, green colour on the plaquettes indicates a $Y$ Pauli term in the operator we measure.}\label{fig:elongated}
\end{figure}

\begin{figure}[hbt!]
    \centering
    \begin{subfigure}{0.45\textwidth}
        \centering
        \includegraphics[width=0.5\textwidth]{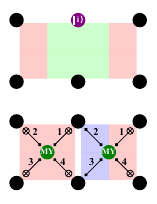}
        \caption{}
        \label{fig:twist_defect1}
    \end{subfigure}
    \hspace{0.05\textwidth}
    \begin{subfigure}{0.45\textwidth}
        \centering
        \includegraphics[width=0.5\textwidth]{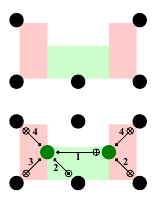}
        \caption{}
        \label{fig:twist_defect2}
    \end{subfigure}
    \caption{Measuring a twist-defect $XXYXX$ stabiliser (a) using our method that requires only a degree-four connectivity device and (b) using the method of \cite{ChamCamtwistbased} that requires a degree-six connectivity device. With our method, we initialise one (accessory) qubit in the middle in the $Y$ basis (in purple), then measure the product $XXXXII\cdot IIZZXX = -XXYYXX$ of the two plaquettes by using schedules that do not satisfy condition (b). The classically flipped outcome of this measurement is then the outcome corresponding to the twist defect.}\label{fig:twist_defect}
\end{figure}

To enable arbitrary logical $X$-$Z$ Pauli measurements, we can use elongated rectangles effectively to change the background lattice, thereby enabling arbitrary twist-free lattice surgery. An example of an elongated rectangle is shown in \Cref{fig:elongated}, where we have six data qubits in a $2\times 3$ arrangement on which we measure a weight-four Pauli product supported on the leftmost two and rightmost two qubits.
The other type of long-range stabilisers, twist defects, are weight-five stabilisers that enable the merging of a patch with others along both its $X$ and $Z$ logical sides, thereby enabling $Y$ measurement on the patch. Twist defects, together with elongated rectangles, are needed for general twist-based lattice surgery. An example is shown in \Cref{fig:twist_defect}. We have the same $2\times 3$ arrangement of data qubits on which we wish to measure a weight-five Pauli product where the Pauli term in the middle is a $Y$.

In order to measure these long-range stabilisers on a degree-four connectivity device, we start with two commuting local weight-four component operator plaquettes that share two joint data qubits. On each joint qubit, one component operator has $X$ Pauli terms, while the other has $Z$. For instance, in \Cref{fig:elongated1,fig:twist_defect1}, these are an $XXXX$ and a $ZZXX$ plaquette. We also initialise in the $Y$ basis those data qubits in the middle on which our stabiliser is not supported; we have two of these for elongated rectangles, and one for twist defects. We shall call these qubits \emph{accessory qubits}. Next, we schedule the syndrome extraction circuits of the two plaquettes in a way that violates condition (b), thereby tangling them. Furthermore, at the last layer of their combined circuit $\mathcal{C}$, instead of measuring the auxiliary qubits in $X$ bases, we measure them in $Y$ bases, resulting in the circuit $\tilde{\mathcal{C}}$. It follows from \Cref{thm:main} that the sum of these two outcomes is the measurement outcome corresponding to the weight-six product, in our example $-XXYYXX$. Since we initialised the accessory qubit(s) in $Y$ basis, we effectively measured the weight-four elongated rectangle $XXIIXX$ or the weight-five twist defect $XXIYXX$. 

Note that, after one round, we require a Clifford correction $\frac{1}{\sqrt{2}}(1-ig_1)^\dagger g_1^{m_1}$, where $g_1=XXXXII$. Therefore, without physically applying the correction, the weight-one stabiliser $IIYIII$ on the accessory qubit becomes a weight-four stabiliser 
\begin{align}
    & \frac{1}{2} g_1^{m_1} (1-ig_1) \cdot IIYIII \cdot (1+ig_1)g_1^{m_1} \nonumber \\ 
    & = \frac{1}{2}(XXXXII)^{m_1}(1-iXXXXII) \cdot IIYIII \cdot (1+iXXXXII)(XXXXII)^{m_1} \nonumber \\ 
    & = (-1)^{m_1}XXZXII.
\end{align}
Thus, at this point, measuring any accessory qubit would destroy the post-measurement state. However, if we do another round of tangled syndrome extraction, we have only a Pauli correction $g_1^{m_1+n_1+1}$ and so the weight-one stabiliser $IIYIII$ becomes 
\begin{equation}
    (-1)^{m_1+n_1+1}IIYIII.
\end{equation}
Therefore, measuring the accessory qubit in the $Y$ basis now does not harm the post-measurement state corresponding to the elongated rectangle or twist defect. Moreover, this outcome is deterministic, $m_1+n_1+1$, under noiseless execution, and so can be used for error detection during syndrome extraction. This additional deterministic measurement outcome can be used in decoding and therefore provides extra information about where errors may have occurred; more details can be found in \Cref{app:detectors}.


\section{Numerical results}\label{sec:numerics}

In this section, we compare the QEC performances of two types of rotated planar code patches: the default patches where each stabiliser is local, and their tangled versions where some stabilisers are replaced by elongated rectangles. We first present our results for quantum memory and then for the stability experiment \cite{Gidney2022stability}. Recall that, in quantum memory, the minimum-weight logical error is space-like (composed of errors on qubits), while in the stability experiment it is time-like (composed of errors that cause measurement failures). Hence, the stability experiment can be used as a prototype to estimate the logical failure rate in an experiment where time-like errors are an additional source of logical failure, such as in lattice surgery or patch-moving. 

For each experiment, we constructed a circuit in terms of the following gates: $Z$-basis reset, $Z$-basis measurement, controlled-$Z$ gate: $CZ$, Hadamard gate: $H$, and a variant of the Hadamard gate that swaps the $Y$ and $Z$ eigenstates (instead of $X$ and $Z$): $H_{YZ}$. These circuits can be found in \cite{gyorgy_pal_geher_2023_8391674}. The noise model we use is parametrised by $p$, the physical error rate, and we apply the following noise operations:
\begin{itemize}
    \item two-qubit depolarising channel with strength $p$ after each $CZ$ gate,
    \item each measurement outcome is flipped classically with probability $p$,
    \item one-qubit gates, reset and measurement are each followed by one-qubit depolarising channels with strength $p/10$,
    \item on each idling qubit in each layer, we apply a one-qubit depolarising channel with strength $p/10$.
\end{itemize}
This noise model captures the idea that, at least for superconducting hardware \cite{Paler2022PipelinedCM}, the most noisy operations are measurements and two-qubit gates.

We used \texttt{stim} \cite{stim} to obtain samples from the noisy circuits and to construct a representation of the decoding (hyper-)graphs. From the samples, we estimated the logical failure probability $p_\ell$ of each experiment, for which we used at least $10^{3s/2}$ number of shots where $s = -\log_{10}p_\ell$. For decoding, in the default cases, we used the minimum-weight perfect matching (MWPM) Python library \texttt{pymatching} \cite{pymatching}. However, we found that, for the tangled version, it is not possible to define an efficient decoding graph in general. Instead, we mapped the circuit-level noise model to a Tanner graph (see \cite{BeliefMatching} for details) and used a more general decoder, \texttt{BPOSD} \cite{bposd, bposdcode}. In the case of quantum memory, we fixed the physical error rate to be $p=10^{-3}$ and performed syndrome extraction for $2$ rounds. Note that $2$ is the minimum number of rounds in the tangled case due to the Clifford nature of the correction after odd numbers of rounds. In the case of stability, we ran experiments on three patches for $4$, $6$ and $8$ rounds, and varied the physical error rate between $p=1.767 \times 10^{-3}$ and $10^{-2}$.


\subsection{Numerical comparison of default and tangled quantum memory simulations}\label{subsec:rotated_qmem}

\begin{figure}[h]
    \centering
    \begin{subfigure}[b]{0.45\textwidth}
        \centering
        \includegraphics[width=0.5\textwidth]{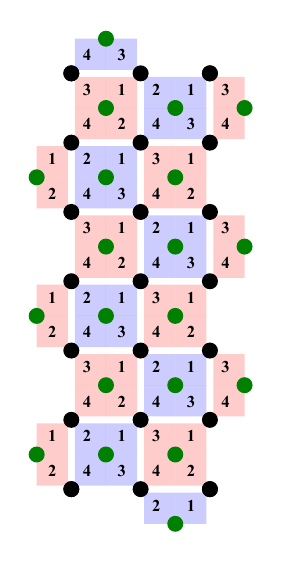}
        \caption{Default rotated planar code patch with $w=3$, $h=7$. The effective $X$- and $Z$-distances are $3$ and $7$, respectively.}
        \label{subfig:def_patch_3x7}
    \end{subfigure}
    \hfill
    \begin{subfigure}[b]{0.45\textwidth}
        \centering
        \includegraphics[width=0.7\textwidth]{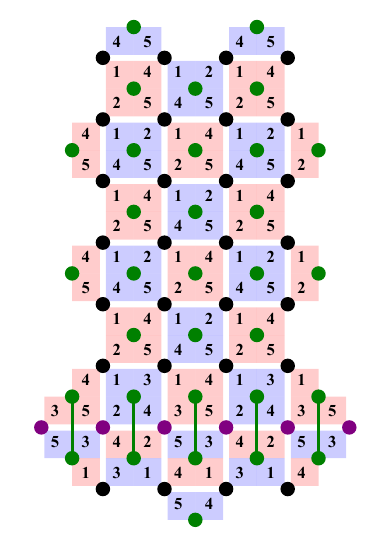}
        \caption{Tangled rotated planar code with $w=4$, $h=7$. The effective $X$- and $Z$-distances are $3$ and $7$, respectively.}
        \label{subfig:tng_patch_4x7}
    \end{subfigure}
    \hfill
    \begin{subfigure}[b]{0.45\textwidth}
        \centering
        \includegraphics[width=0.8\textwidth]{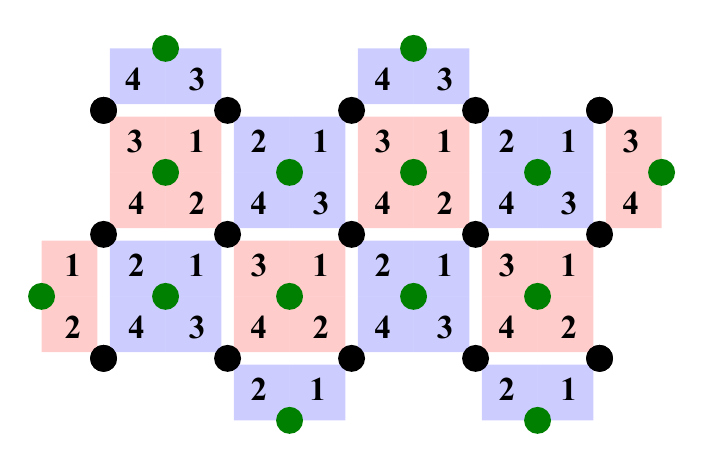}
        \caption{Default rotated planar code patch with $w=5$, $h=3$. The effective $X$- and $Z$-distances are $5$ and $3$, respectively.}
        \label{subfig:def_patch_5x3}
    \end{subfigure}    
    \hfill
    \begin{subfigure}[b]{0.45\textwidth}
        \centering
        \includegraphics[width=1\textwidth]{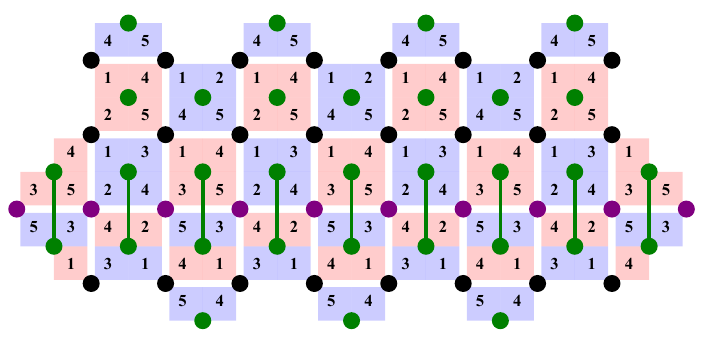}
        \caption{Tangled rotated planar code with $w=8$, $h=3$. The effective $X$- and $Z$-distances are $5$ and $3$, respectively.}
        \label{subfig:tng_patch_8x3}
    \end{subfigure}   
    \caption{Default and tangled versions of two rotated planar code patches. In (b) and (d), the green lines between auxiliary qubits indicate a tangled schedule between the two corresponding component plaquettes. Furthermore, the accessory qubits are shown in purple.}\label{fig:rotated_patch_sizing}
\end{figure}

A default rotated planar code depends on two parameters, namely the minimal numbers of $X$- and $Z$-type data qubit errors that lead to a logical failure, which we call the $X$- and $Z$-distances, respectively. We choose to place the planar code in such a way that the horizontal edges of the patch are of $Z$ type, and hence the logical $Z$ operator is vertical; see \Cref{subfig:def_patch_3x7} and \Cref{subfig:def_patch_5x3}. Therefore, the height $h$ of a patch is equal to its $Z$-distance $d_z^{def}$, while its width $w$ coincides with its $X$-distance $d_x^{def}$. To extract the syndromes for the default rotated planar code, we use the usual $N$- and $Z$-shaped schedules with four layers of entangling gates. More precisely, the $N$-shaped schedule is that used for red plaquettes in e.g. \Cref{subfig:def_patch_3x7}, while the $Z$-shaped schedule is that used for blue plaquettes. Note that the minimum numbers of fault locations that lead to a logical error during $X$- and $Z$-quantum memory using a particular set of schedules are called the effective $Z$- and $X$-distances, respectively. It is well-known that, with the $N$-$Z$-shaped schedules, the default patch has the effective distances $h$ and $w$; see e.g. \cite{Tomita_2014}.

Two examples of tangled rotated planar codes are shown in \Cref{subfig:tng_patch_4x7,subfig:tng_patch_8x3}. Note that, even though we use one additional row of qubits for the tangled case, we still use the same number of stabilisers in the vertical direction and hence we shall say that the height of the tangled patch is $h$ instead of $h+1$. The scheduling we use for the tangled version is also shown. The reader may notice that we apply four entangling gates on data and accessory qubits in the bulk; however, we apply entangling gates in five layers instead of four. Indeed, after an extensive search, we concluded that a scheduling for the tangled version with four entangling layers that does not propagate $X$/$Z$-type errors from the auxiliary qubits of regular stabilisers parallel to the corresponding logical operator's direction does not exist. The schedules in the tangled case (\Cref{subfig:tng_patch_4x7,subfig:tng_patch_8x3}) follow a similar pattern to the default version for regular stabilisers, and we use two types of $N$-shaped schedules for the component operators. In fact, after some initial simulation we found that the particular scheduling of component operators has little effect on the effective distance or the QEC performance of the tangled patch. However, the scheduling of \Cref{subfig:tng_patch_4x7,subfig:tng_patch_8x3} is convenient as it extends very naturally below the bottom boundary (see \Cref{app:tangled_middle}) and hence is applicable for more general patches too (e.g. in lattice surgery). Furthermore, we point out that, in each round of syndrome extraction, each qubit is acted on with at most four entangling gates. Hence the five layers of entangling gates only introduce one additional idling layer on the qubits, which is less harmful than additional two-qubit gates would be. 

For the effective distances of the tangled patches with width $w$ and height $h$, we find the following formulae, on which we give more detail in \Cref{app:rotated-tangled-logical-errors}:
\begin{equation}
    d_z^{tng} = h \quad\text{and}\quad d_x^{tng} = \left\lfloor\frac{w}{2}\right\rfloor+1.
\end{equation}
For instance, the patches in \Cref{subfig:def_patch_3x7} and \Cref{subfig:tng_patch_4x7} have the same effective distances, as do the patches in \Cref{subfig:def_patch_5x3} and \Cref{subfig:tng_patch_8x3}. Intuitively, this is approximately what we would expect, since each error on an auxiliary qubit can propagate into an at most weight-two error on data qubits that are not accessory qubits. Moreover, if this propagated error is of $ZZ$ type, then its direction is always horizontal, i.e. perpendicular to the logical $Z$ direction. Note that this halving of one type of effective distance ($X$ in this case) is typical for other syndrome-extraction techniques as well that use more than one auxiliary qubit. For instance, in \Cref{fig:elongated2}, an error from an auxiliary qubit could spread to a weight-two $XX$ vertical data qubit error, regardless of the direction of the logical $X$ operator. Of course, we need to compensate for this effect somehow and, in the next section and in \Cref{app:rotated_lattice_surgery}, we will demonstrate how to do that for lattice surgery without increasing the number of qubits of the QPU.

In our simulations, we compared the default $h\times w$ rotated planar code's $X$/$Z$-memory performance with the tangled version's for size $h\times (2w-2)$. Since their effective distances agree, we expected to see comparable results. We fixed the physical error rate to be $p=10^{-3}$, the number of rounds to be $2$, and compared the logical failure rates. We present our numerical results below for three cases: the narrow case when $w=3$ is fixed while $h$ varies, the wide case when $h=3$ is fixed while $w$ varies, and the squarerer case when we vary $h$ and set $w=h$.

\begin{figure}[t]
    \centering
    \begin{subfigure}[b]{0.48\textwidth}
    \includegraphics[width=\textwidth]{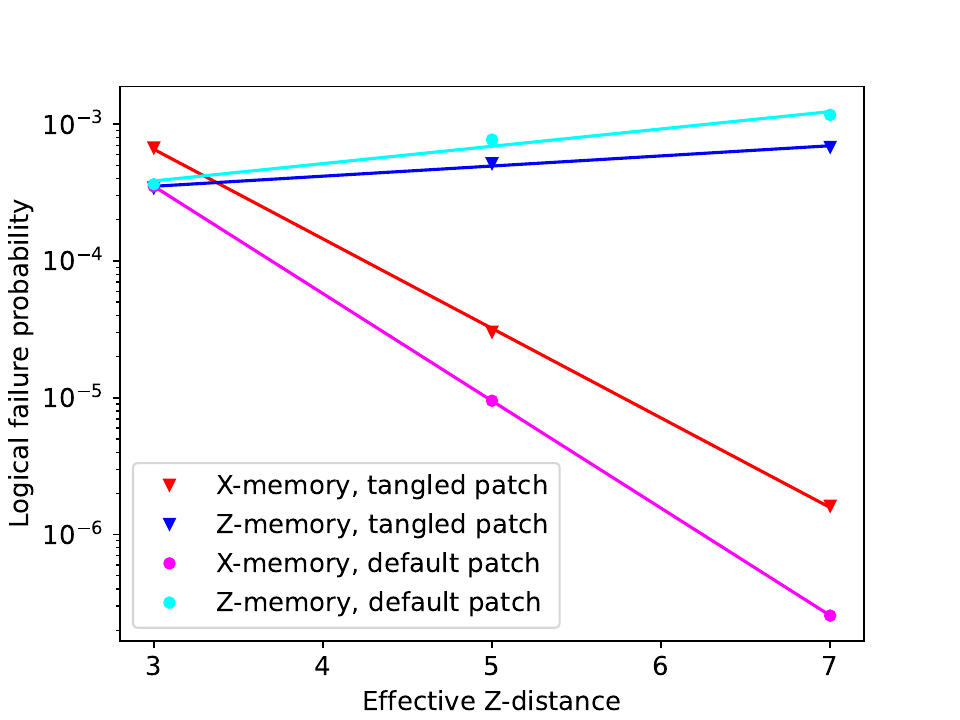}
    \caption{}\label{fig:narrow_data}
    \end{subfigure}
    \begin{subfigure}[b]{0.48\textwidth}
    \includegraphics[width=\textwidth]{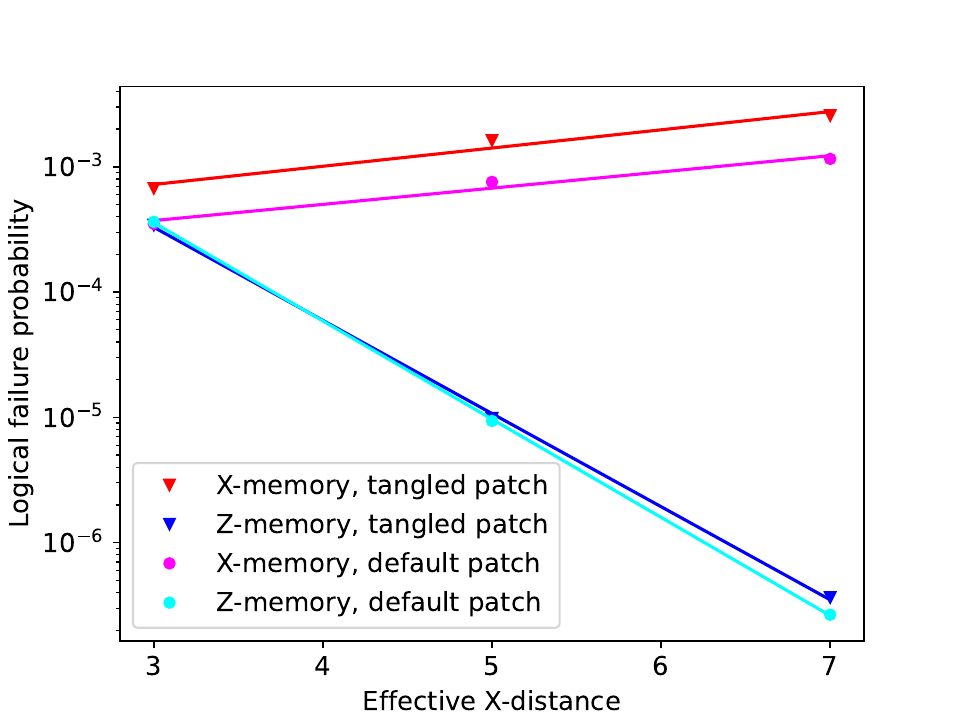}
    \caption{}\label{fig:wide_data}
    \end{subfigure}
    \begin{subfigure}[b]{0.48\textwidth}
    \includegraphics[width=\textwidth]{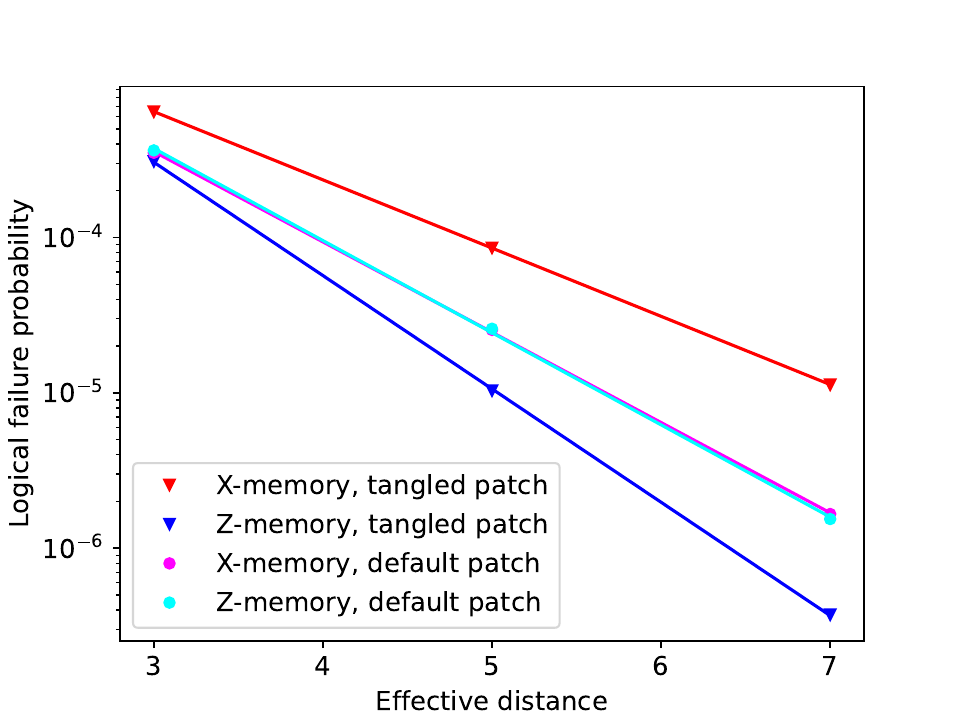}
    \caption{}\label{fig:squarerer_data}
    \end{subfigure}
    \caption{Comparing the default and tangled patches' quantum memory performance. (a) The narrow case where the effective $X$ distance, $w$, is fixed to be $3$ and the physical error rate is $p=10^{-3}$. This means that the width is fixed to be $3$ for default and $4$ for tangled planar codes, while the height varies as shown on the horizontal axis. (b) The wide case when the effective $Z$ distance, $h$, is fixed to be $3$, the effective $X$ distance, $w$, varies as shown on the horizontal axis, and the physical error rate is $p=10^{-3}$. This means that the height is fixed to be $3$ for both default and tangled planar codes, while the width is $w$ for the default and $2w-2$ for the tangled case. (c) The squarer-shaped case when the joint effective distance, $h=w$, varies as shown on the horizontal axis. This means that the default patch is of size $h\times h$, while the tangled patch is of size $(2h-2)\times h$.}\label{fig:narrow_and_wide_data}
\end{figure}

Our results are shown in \Cref{fig:narrow_and_wide_data}. For the narrow patch case (\Cref{fig:narrow_data}), the effective $Z$-distance varies as $h = 3, 5, 7$. We performed a log-linear fit (with base $e$) for both the default and the tangled versions, and found the following gradients: $-1.8063$ for default $X$-memory, $-1.5075$ for tangled $X$-memory, $0.2918$ for default $Z$-memory, and $0.1707$ for tangled $Z$-memory. Note that we expect the logical failure rate to decrease exponentially as $h$ grows for the case of $X$ memory, and increase in the case of $Z$ memory. Therefore, under a physical error rate of $p=10^{-3}$, these gradients mean the following: in order to match the logical failure probability for $X$ quantum memory experiment of a default $h\times 3$ patch, we need to choose the effective $Z$-distance of the tangled patch approximately to be $e^{-1.5075+1.8063} \cdot h = 1.3482 \cdot h$; while for the case of $Z$ quantum memory the effective $Z$-distance has to be $e^{0.1707-0.2918}\cdot h = 0.8859\cdot h$ in the tangled case.

For the wide patch case (\Cref{fig:wide_data}) where the effective $X$-distance varies as $w = 3, 5, 7$, we performed a similar log-linear fit as in the narrow case, and found the following gradients: $0.2980$ for default $X$-memory, $0.3348$ for tangled $X$-memory, $-1.8051$ for default $Z$-memory, and $-1.7116$ for tangled $Z$-memory. As for the narrow case, we can conclude similarly the following under a physical error rate of $p=10^{-3}$: in order to match the logical failure probability of a default $3\times w$ patch, in the case of $X$ quantum memory we need to use a tangled patch with effective $X$-distance $1.0374\cdot w$; while in the case of $Z$ memory this effective $X$-distance needs to be $1.0980\cdot w$ for the tangled patch. 

For the squarer patch case (\Cref{fig:squarerer_data}) where the effective $X$- and $Z$-distances are equal, i.e. $h=w$, and this parameter varies as $w = 3, 5, 7$, we again performed a log-linear fit. We found the following gradients for this case: $-1.3384$ for default $X$-memory, $-1.0114$ for tangled $X$-memory, $-1.3660$ for default $Z$-memory, and $-1.6800$ for tangled $Z$-memory. Similarly to the above two cases, we can conclude that under a physical error rate of $p=10^{-3}$: in order to match the logical failure probability of a default $h\times h$ patch, in the case of $X$ quantum memory we need to use a tangled patch with effective distance $1.3868\cdot h$; while in the case of $Z$ memory this effective distance needs to be $0.7305\cdot w$ for the tangled patch.

We expect the multiplicative factors to decrease when we decrease the physical error rate $p$ and/or increase the parameter ($h$ for the wide, or $w$ for the narrow case) of the patch that was fixed here. From these results we conclude that the quantum memory performances of default and tangled cases are approximately the same, as long as the effective distances are matched. Finally, we note that in order to symmetrise the $X$- and $Z$-memory performances in the squarer case, we may decrease the width of the tangled patch by $1$ or $2$. Then, on the one hand, the qubit count is less and the effective $X$-distance decreases by $1$ but, on the other hand, there are fewer minimal-length vertical $Z$-logical strings present. Therefore, we expect the $X$-memory performance to improve and the $Z$-memory performance to get somewhat worse. In this way, we would expect that, for large patches, one could match and symmetrise the logical performance of the tangled case with that of the default case more closely by using fewer qubits, even though the effective distances are not exactly matched.


\subsection{Numerical comparison of default and tangled stability experiments}\label{subsec:stability}

\begin{figure}[h]
    \begin{subfigure}[b]{0.3\textwidth}
        \centering
        \includegraphics[width=0.8\textwidth]{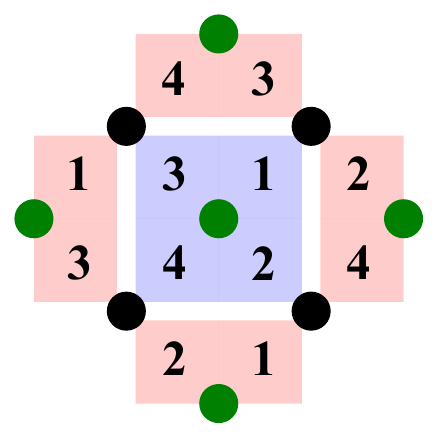}
        \caption{Default $2\times 2$ rotated planar patch for the stability experiment}
        \label{subfig:def_stab_patch_2x2}
    \end{subfigure}
    \hfill
    \begin{subfigure}[b]{0.3\textwidth}
        \centering
        \includegraphics[width=0.8\textwidth]{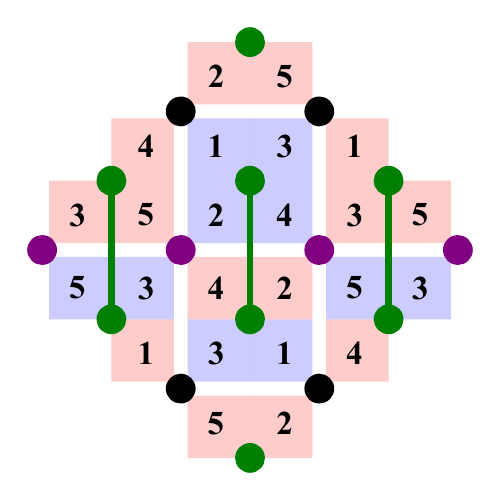}
        \caption{Tangled $2\times 2$ rotated planar patch for the stability experiment}
        \label{subfig:tng_stab_patch_2x2}
    \end{subfigure}
    \hfill
    \begin{subfigure}[b]{0.3\textwidth}
        \centering
        \includegraphics[width=0.7\textwidth]{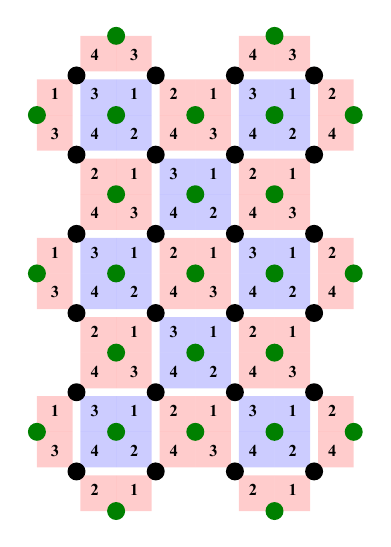}
        \caption{Default $4\times 6$ rotated planar patch for the stability experiment}
        \label{subfig:def_stab_patch_4x6}
    \end{subfigure}
    \hfill
    \begin{subfigure}[b]{0.45\textwidth}
        \centering
        \includegraphics[width=0.7\textwidth]{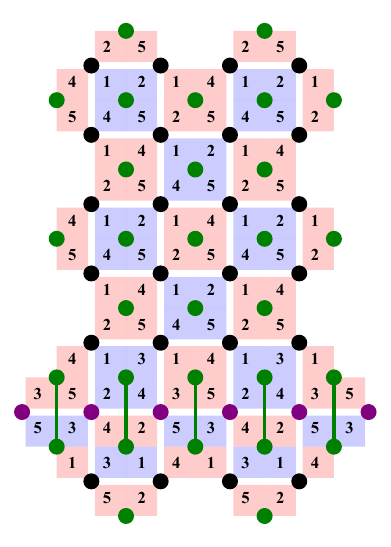}
        \caption{Tangled $4\times 6$ rotated planar patch for the stability experiment}
        \label{subfig:tng_stab_patch_4x6}
    \end{subfigure}
    \hfill
    \begin{subfigure}[b]{0.45\textwidth}
        \centering
        \includegraphics[width=1\textwidth]{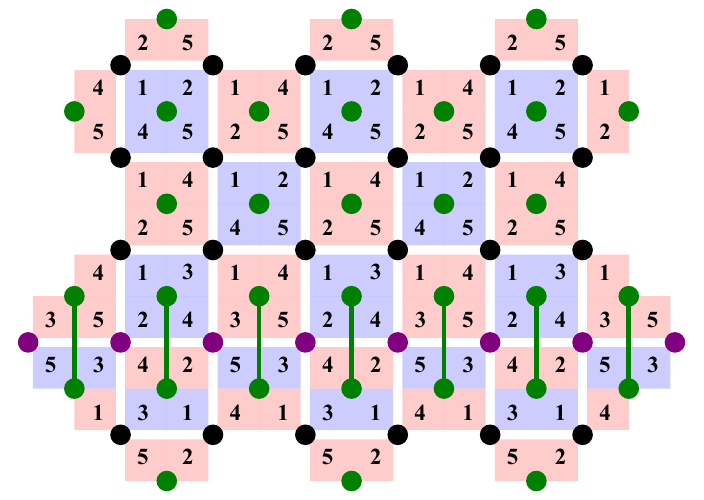}
        \caption{Tangled $6\times 4$ rotated planar patch for the stability experiment}
        \label{subfig:tng_stab_patch_6x4}
    \end{subfigure}
    \hfill
    \caption{Default and tangled rotated planar code patches without logical corners. The $X$-type stabilisers are over-determined, hence their measurement outcomes sum to $0$ (mod $2$).}\label{fig:stability_patches}
\end{figure}

The stability experiment \cite{Gidney2022stability} was proposed to test for time-like failures, and Gidney calls it the dual of a quantum memory experiment. In this subsection, we compare the stability experiment performances of the default and tangled versions of rotated planar code patches that have no logical corners; see \Cref{fig:stability_patches}.  We consider a planar patch where one type of stabilisers, say the $X$ type, is over-determined, i.e. they multiply into the identity operator. The stability experiment proceeds as follows: initialise all data qubits in the $Z$ basis; perform $n$ rounds of syndrome extraction; and finally measure out all data qubits in the $Z$ basis. In the absence of errors, for any round, the product of the $X$-type stabiliser outcomes has zero parity, and this is our logical observable (say from the first round) we use to validate our experiment after error correction. At least $n$ faults (e.g. $n$ measurement errors on the same $X$-type stabiliser) are required for an undetectable logical failure \cite{Gidney2022stability}, hence the experiment's effective distance is $n$. The stability experiment can be used as a proxy to estimate the number of rounds needed for lattice surgery in the merge stage, or in patch movement for the growing step, or essentially in any experiment where a time-like failure is the main cause of logical failure.

We chose three tangled patches on which we performed stability experiments: a $2\times 2$, a $4\times 6$ and a $6\times 4$ patch; see \Cref{subfig:tng_stab_patch_2x2,subfig:tng_stab_patch_4x6,subfig:tng_stab_patch_6x4}. We compared their stability performances to the stability performance of default patches that have the same number of $X$- and $Z$-type stabilisers; see \Cref{subfig:def_stab_patch_2x2,subfig:def_stab_patch_4x6}. Note that the particular scheduling used on default patches has essentially no effect on the stability experiment performance \cite{Gidney2022stability}, hence we compare the two larger tangled patches to the same default patch.

\begin{figure}[h]
    \centering
    \includegraphics[width=0.45\textwidth]{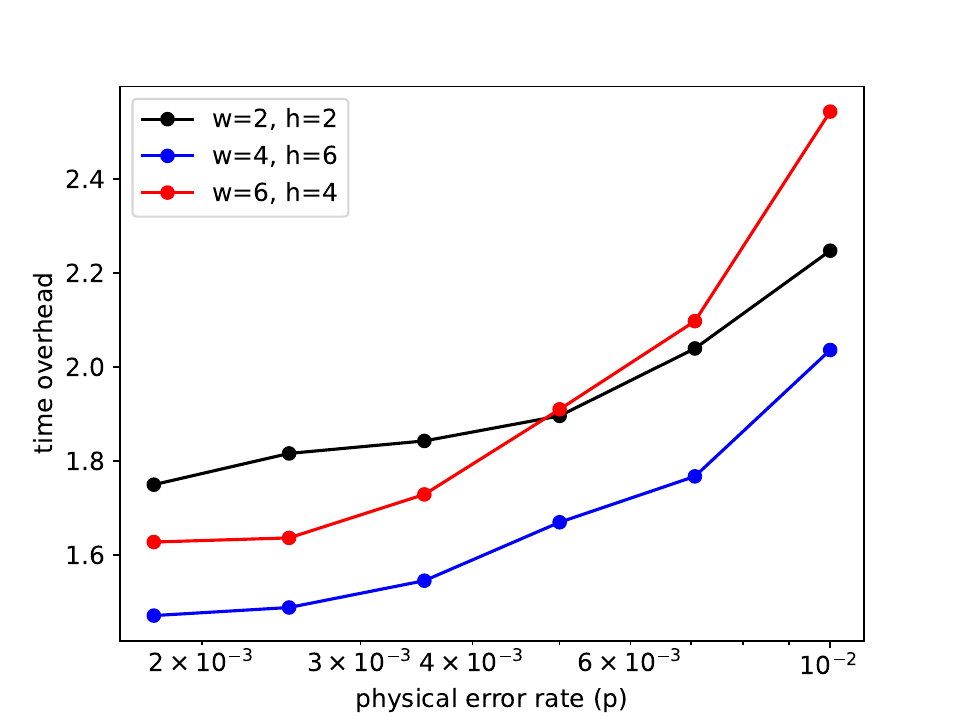}
    \caption{Comparing the default and tangled stability experiments for the patches of \Cref{fig:stability_patches}. The time overhead $R=\gamma^{tng}/\gamma^{def}$ is plotted against the physical error rate. }\label{fig:stab_time_overhead}
\end{figure}

For each patch, we simulated a stability experiment with $n=4, 6, 8$ rounds. We varied the physical error rate between $1.767 \times 10^{-3}$ and $10^{-2}$. In \Cref{fig:stab_time_overhead}, we show the \emph{time overhead}, $R = \gamma^{tng}/\gamma^{def}$, where the two constants ($a$ and $\gamma$) are obtained from the line-fittings 
\begin{equation}
    \log (p_l^{def}) = \log (a^{def}) - \gamma^{def} n \quad\text{and}\quad\log (p_l^{tng}) = \log (a^{tng}) - \gamma^{tng} n
\end{equation} 
for the logarithms of the logical failure rates. We interpret the time overhead $R$ as a scalar factor we need to multiply with the default number of rounds $n$ in order to match the stability experiment's performance in the tangled case. \Cref{fig:stab_time_overhead} shows clearly that the time overhead decreases as the physical error rate decreases. Moreover, it is already close to, or less than, $R=2$ at around the quantum memory threshold. Furthermore, if we increase the size of the patch, the logical failure rate improves in the case when $1.767 \times 10^{-3}<p<5\times 10^{-3}$. Therefore, by extrapolating the plotted data from \Cref{fig:stab_time_overhead}, we would expect that for large patches, which we typically need for lattice surgery, the time overhead would satisfy $R<1.5$ for physical error rate $p<1.767 \times 10^{-3}$, with $R$ potentially decreasing further at even lower $p$.

\section{An architecture for fault-tolerant quantum computation with the unrotated planar code}\label{sec:full_lattice_surgery}

In this section, we present a proposal for performing general multi-logical-qubit Pauli measurements on the square-grid connectivity QPU, as in \Cref{fig:planar_qpu}. Recall that this makes it possible to perform full FTQC \cite{GoSc, ChamCamtwistbased, ChamCamtwistfree}. More precisely, we propose a general twist-based lattice surgery scheme using the unrotated planar code. Note that we discuss the case of the rotated planar code in \Cref{app:rotated_lattice_surgery}.

\begin{figure}[hbt!]
    \centering
    \begin{subfigure}[b]{0.49\textwidth}
        \centering
        \includegraphics[width=0.9\textwidth]{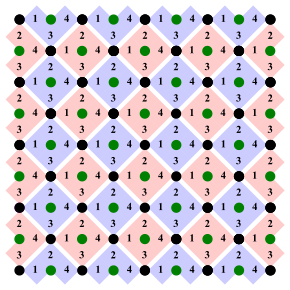}
        \caption{}\label{subfig:5x5_default_unrotated}
    \end{subfigure}
    \hfill
    \begin{subfigure}[b]{0.49\textwidth}
        \centering
        \includegraphics[width=0.9\textwidth]{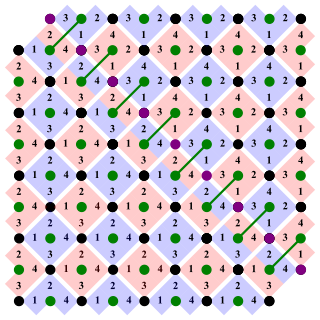}
        \caption{}\label{subfig:5x5_tangled_unrotated}
    \end{subfigure}
    \caption{Default (a) and tangled (b) versions of a $5\times 5$ unrotated planar code. Their scheduling is also shown.}\label{fig:unrotated_scheduling}
\end{figure}

Regarding which version of the planar code to base an architecture on, Beverland \textit{et. al.} \cite{Beverland_2019} showed that the rotated and unrotated toric codes have very similar logical failure rates for the same number of qubits under a phenomenological noise model. Recently, circuit-level noise simulations \cite{Paler2022PipelinedCM} for both types of planar codes have also confirmed only marginal differences in the efficiency of these codes. Therefore, even though the rotated planar code has gained popularity, there is no evidence that it offers a substantial advantage for large-scale architectures. Furthermore, we point out that a default unrotated planar code patch has the desirable property that its stabilisers can be scheduled in a uniform manner (\Cref{subfig:5x5_default_unrotated}), unlike in the rotated case. This is because there is no damage to the effective distance if we spread errors from auxiliary qubits in any diagonal direction. This is useful for our tangled schedules, as it becomes possible to design a syndrome extraction circuit with $4$ entangling layers per QEC round, as shown in \Cref{subfig:5x5_tangled_unrotated}, instead of $5$ in the rotated planar examples in \Cref{sec:numerics}. Due to less idling on qubits, we would expect slightly improved QEC performance in this case. Therefore, here, we choose to present an architecture using the unrotated code, since we found this especially amenable to using tangled schedules.

\begin{figure}[hbt!]
    \centering
    \includegraphics[width=0.85\textwidth]{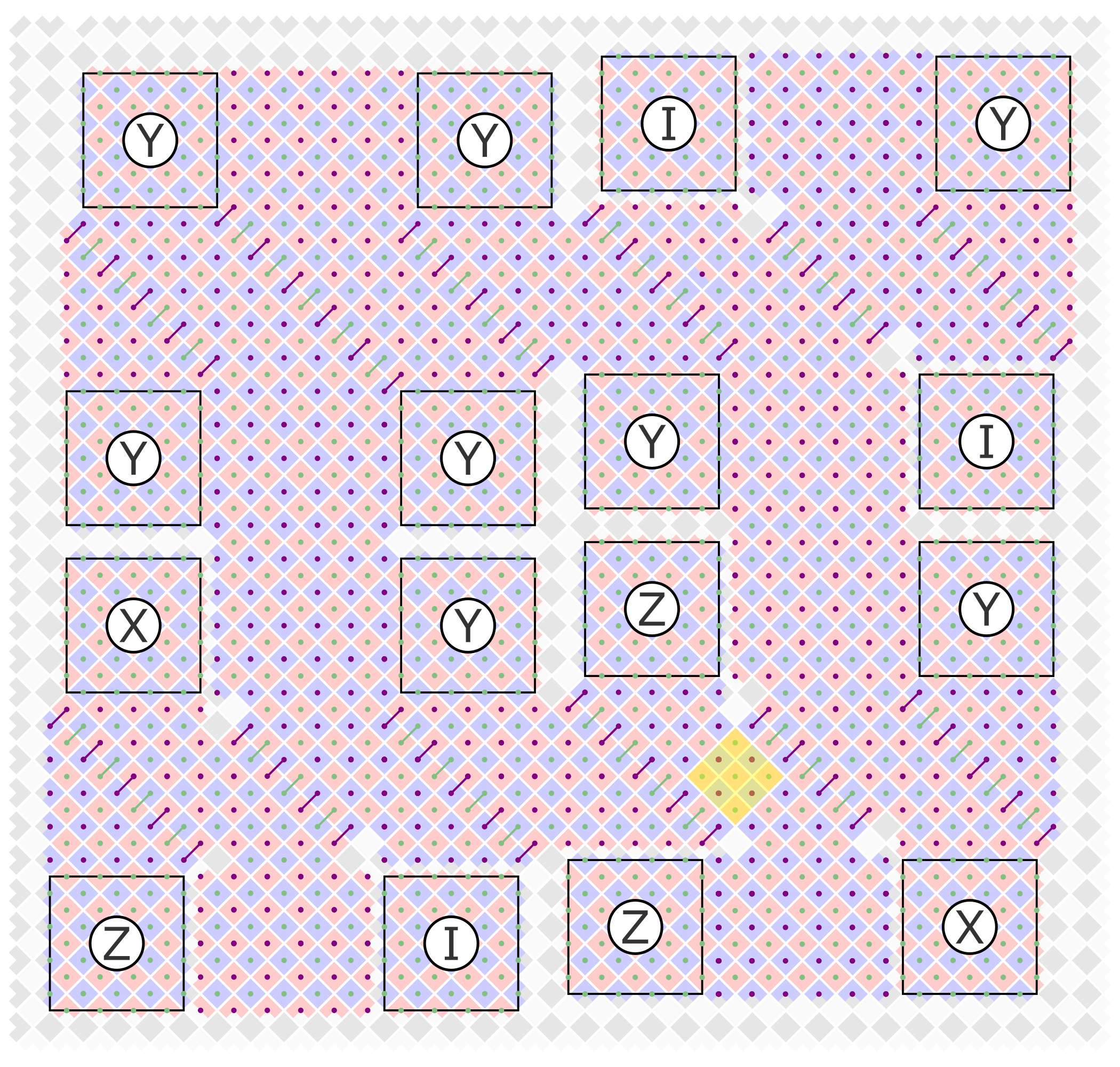}
    \caption{An example showing the merge-stage patch of a general lattice surgery operation involving several logical qubits as unrotated planar code patches. The original patches are outlined in black and labelled with their Pauli terms in the measurement. Here and in subsequent plots, the purple dots and lines label those outcomes that multiply to give the logical Pauli product we measure. Note that, at certain places, we could shrink the merge-stage patch (e.g. at the bottom-left, we do not need the rectangle-shaped extension) if we do not wish to involve further patches in our logical measurement. However, we left these extensions here to show how other patches from adjacent tiles could be merged in easily too. Also, we could omit some of the merging through the boundaries of tile units, e.g. in the present case we could simply join at three boundaries instead of four. The $(d-1)\times(d-1)$ sized yellow square is an area we need to cross in order to have a logical failure string that connects the yellow square's top and bottom vertices, thus connecting two boundaries of the merge-stage patch.}\label{fig:unrotated_general_lattice_surgery}
\end{figure}

Inspired by the architecture presented in \cite[Fig. 11 and 14]{ChamCamtwistfree}, we tile the QPU in the bulk with units that could accommodate approximately $9$ logical patches of the same square size, of which $5$ are used for routing. We allocate the logical patches in the bulk as depicted in \Cref{fig:unrotated_general_lattice_surgery}, where four tile units are depicted and which also shows an example of a concrete merge-stage patch for twist-based lattice surgery. In \Cref{app:unrotated_cut_inside}, we give more details on how to construct the merge-stage patch for a general lattice surgery operation. We find that the effective distance of such a general merge-stage patch is always at least $d-1$, given that the logical patches have distance $d$. Below, we show this for the example of \Cref{fig:unrotated_general_lattice_surgery}, but the general case can be proven similarly.

\begin{figure}[hbt!]
    \centering
    \begin{subfigure}[b]{0.32\textwidth}
        \centering
        \includegraphics[width=\textwidth]{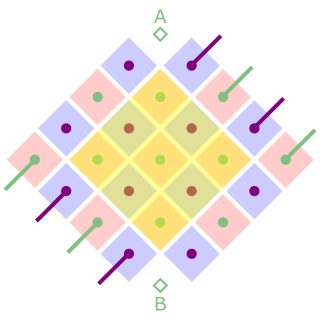}
        \caption{}\label{subfig:yellow_area}
    \end{subfigure}
    \hfill
    \begin{subfigure}[b]{0.32\textwidth}
        \centering
        \includegraphics[width=\textwidth]{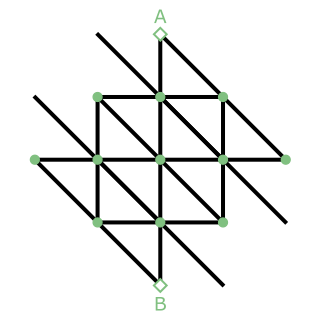}
        \caption{}\label{subfig:yellow_area_eff_dist_graph_1}
    \end{subfigure}
    \hfill
    \begin{subfigure}[b]{0.32\textwidth}
        \centering
        \includegraphics[width=\textwidth]{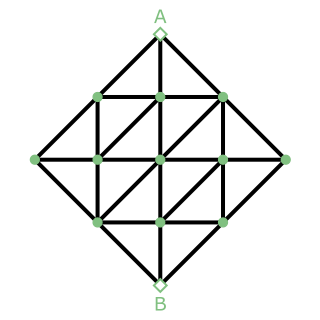}
        \caption{}\label{subfig:yellow_area_eff_dist_graph_2}
    \end{subfigure}
    \caption{(a) The area including the yellow square from \Cref{fig:unrotated_general_lattice_surgery}. The elongated stabilisers always spread the errors from their auxiliary qubits from the north-west to the south-east. For the regular $Z$-stabilisers we consider two cases.
    (b) If regular $Z$-stabilisers also spread errors from their auxiliary qubits in the north-west to the south-east direction (i.e. as the elongated ones), we can compose the shown graph where each edge corresponds to a single fault location during syndrome extraction, and as a result flips the outcome of the stabilisers that correspond to the vertices of that edge. Vertices $A$ and $B$ are boundary nodes. A logical error is a path in the graph that connects $A$ and $B$. This shows that the effective distance of the merge-stage patch in \Cref{fig:unrotated_general_lattice_surgery} is $d-1$.   
    (c) If regular $Z$-stabilisers spread errors in the south-west to north-east direction (i.e. perpendicular to the direction of spreading of the elongated stabilisers), e.g. as is the case in \Cref{subfig:5x5_tangled_unrotated}, then this graph can be composed. Again, it is straightforward to see that the effective distance is $d-1$.
    }
\label{fig:eff-distance-with-graph}
\end{figure}

We consider the effective distance, which depends on our schedule choice. If we choose the schedules in one of the usual ways for regular stabilisers of the same type, namely spreading errors from auxiliary qubits to data qubits in diagonal directions, then the effective distance of a general merge-stage patch is $d-1$ in the worst-case scenario. Indeed, we only need to be careful along a diagonal array of elongated rectangles and twist defects, and in between two neighbouring such diagonal arrays. Note that the elongated stabilisers in such a diagonal array spread the errors from their auxiliary qubits in the north-west to the south-east direction. Therefore, along any diagonal array we need at least $l/2$ failures to have a logical error string, where $l$ is the number of qubits along the long side of the array. However, this length is always at least $l=2(d-1)$ in our case. 

As for the case in between two neighbouring diagonal arrays of elongated rectangles and twist defects, let us draw a yellow square that contains $(d-1)\times (d-1)$ data qubits, as shown in \Cref{fig:unrotated_general_lattice_surgery}. This area is depicted in \Cref{subfig:yellow_area} for clarity. We have two cases depending on the scheduling of regular stabilisers. First, assume that regular stabilisers of the relevant type spread errors from the auxiliary qubits in the north-west to the south-east direction (i.e. the same as the elongated rectangles). We observe that a logical failure string connecting one boundary of the patch to the other has to pass through this yellow square connecting its top vertex to its bottom vertex, which are indicated as boundary vertices $A$ and $B$ in \Cref{fig:eff-distance-with-graph}. More precisely, we can compose the graph \Cref{subfig:yellow_area_eff_dist_graph_1}, where edges correspond to single fault locations and the vertices of each edge to stabilisers whose outcome is flipped due to that fault. The horizontal and vertical edges correspond to single data qubit $Z$ errors, while the north-west to the south-east diagonal edges correspond to auxiliary qubit errors that spread into weight-$2$ data qubit errors. A logical error corresponds to a path in this graph connecting $A$ to $B$. It is clear that such a path has length at least $d-1$, which implies that we need at least $d-1$ fault locations to have a logical failure. Second, suppose that regular stabilisers of the relevant type spread errors in the perpendicular south-west to north-east direction in the yellow area. Then, we can compose a similar graph as shown in \Cref{subfig:yellow_area_eff_dist_graph_2}. It is again straightforward to see that any path connecting $A$ and $B$ has length at least $d-1$ and therefore, the effective distance is at least $d-1$ in this case too.

We have shown that, using tangled syndrome extraction, we can perform general twist-based lattice surgery on degree-four connectivity hardware without adding additional qubits, thereby relaxing the hardware requirement. We are not aware of any previous lattice surgery protocol in the literature that provides a method for doing this without adding significant qubit-overhead. For instance, the protocol of \cite[Fig. 4]{ChamCamtwistbased} uses degree 6 connectivity (and may require degree 8 in some situations).  

However, our scheme does incur a time overhead relative to a higher connectivity scheme, as discussed in \Cref{subsec:stability}. This means that there is some increased time-cost to achieve the same logical fidelity; however this decreases as the patch size increases, and as the physical error rate decreases. 

Note also that in the protocol of \cite[Fig. 4]{ChamCamtwistbased} one fault location on an elongated stabiliser's auxiliary qubit can lead to a weight-$2$ data qubit error that aligns with a logical string.  Fortunately, in the context of $Y \otimes Y$ lattice surgery, the relevant logical operator is length $\sim 2d$ and so this is not a problem in practice.  A similar effect is seen with our tangled-schedules approach -- the distance is halved for small demonstrations, but employing the protocols described in this work to perform lattice surgery operations means that this effect is not relevant. 

\section{Conclusions}\label{sec:conclusion}
In this work, we have presented a scheme for full fault-tolerant quantum computation using the surface code on a uniform, square-grid connectivity device. Such connectivity is common in hardware based on, for example, superconducting qubits, where increasing the connectivity is difficult due to crosstalk and engineering challenges. The scheme is based on ``tangling'' the respective syndrome extraction circuits of local, low-weight component operators, enabling the measurement of their non-local and/or high-weight product without increasing the required connectivity or the number of two-qubit gates. As examples, we have shown how the tangled syndrome extraction circuits can be used to measure elongated-rectangle and twist-defect stabilisers, which are required for fault-tolerant computation using lattice surgery. 

Through numerical simulations, we demonstrated the performance of the method in quantum memory and stability experiments. For quantum memory, we found that, if the effective distances are matched, a planar code patch using our tangled syndrome extraction method performs very similarly to a patch using standard syndrome extraction. For stability experiments, we found that we require a time overhead for the patch with tangled syndrome extraction to match that with standard syndrome extraction. However, this overhead is around a factor of $2$ at the quantum memory threshold for even the smallest patch sizes, and we observe that this overhead decreases with decreasing physical error rate or increasing patch size. Therefore, the time overhead is smaller than the $2\times$ overhead of Ref. \cite{ChamCamtwistfree}.

We have also presented a general theorem that can be used to measure the product of an arbitrary set of component operators, provided the tangling structure is tree-like. In this work, we have only considered applications where a stabiliser is a product of two component operators. It is, however, a very natural next step to explore schemes where we use more than two component operators to measure a stabiliser. Furthermore, whilst we have focused on planar codes in this work, we believe our tangled syndrome extraction method has applications for other stabiliser codes, and this would be another natural continuation of the present work.


\section*{Acknowledgments}
We would like to thank the two anonymous referees for their valuable suggestions.

\bibliography{main}

\appendix\label{appendix}

\section{Further technical details}\label{app:a}
All the figures for this and the forthcoming sections can be found at the end of the paper.

In this section of the Appendix, we present some technical details that we did not fully discuss in the main part of the paper. First, we explain why elongated rectangles are needed to do general lattice surgery that does not involve $Y$ Pauli terms. Then, we discuss how to schedule a tangled rotated planar code patch that has the elongated rectangles in the middle. After that, we explain how to define so-called detectors for enhanced error correction for elongated rectangles and twist defects when their syndrome extraction is done via our tangled schedules technique. Then, we discuss the effective distance of tangled patches, and present a method to construct the merge-stage patch for general lattice surgery using the unrotated planar code. We conclude the section with a discussion on some alternative FTQC models that do not involve PBC.

\subsection{Background lattice} \label{app:background-lattice}
\begin{figure}[hbt!]
    \centering
    \begin{subfigure}[b]{\textwidth}
        \centering
        \includegraphics[width=0.49\textwidth]{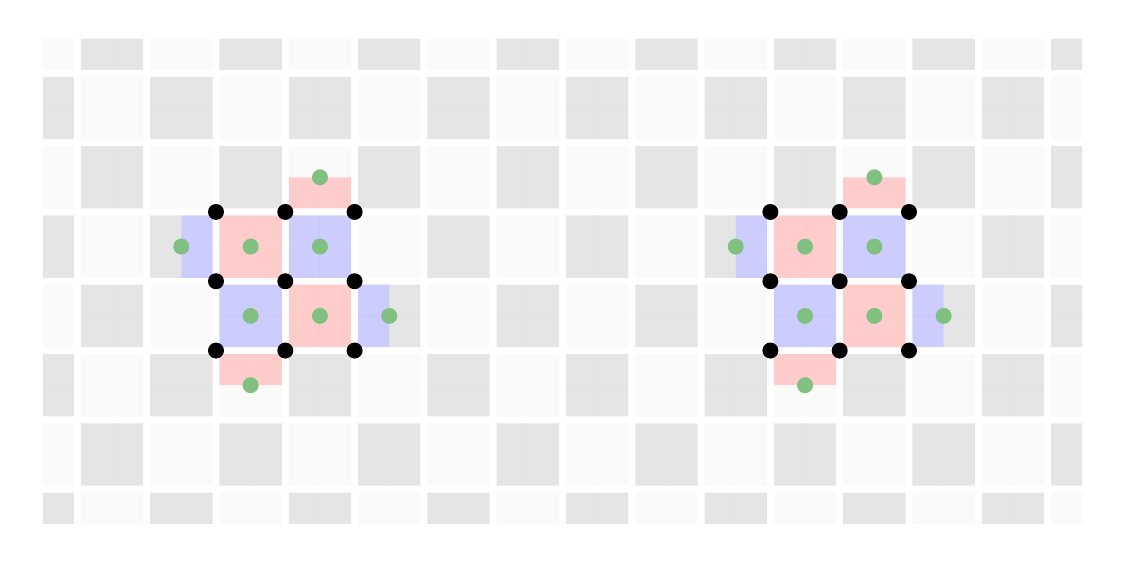}
        \caption{}\label{subfig:need_for_elongated_two_patches}
    \end{subfigure}
    \begin{subfigure}[b]{0.49\textwidth}
        \centering
        \includegraphics[width=\textwidth]{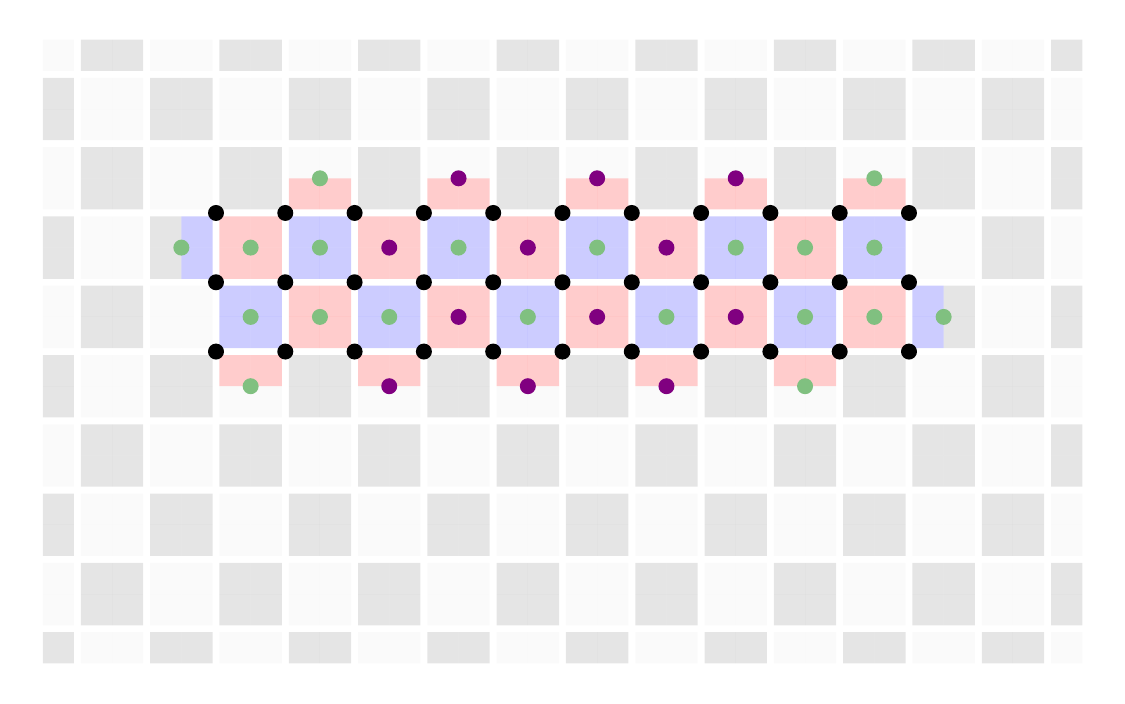}
        \caption{}\label{subfig:need_for_elongated_two_patches_xx}
    \end{subfigure}
    \hfill
    \begin{subfigure}[b]{0.49\textwidth}
        \centering
        \includegraphics[width=\textwidth]{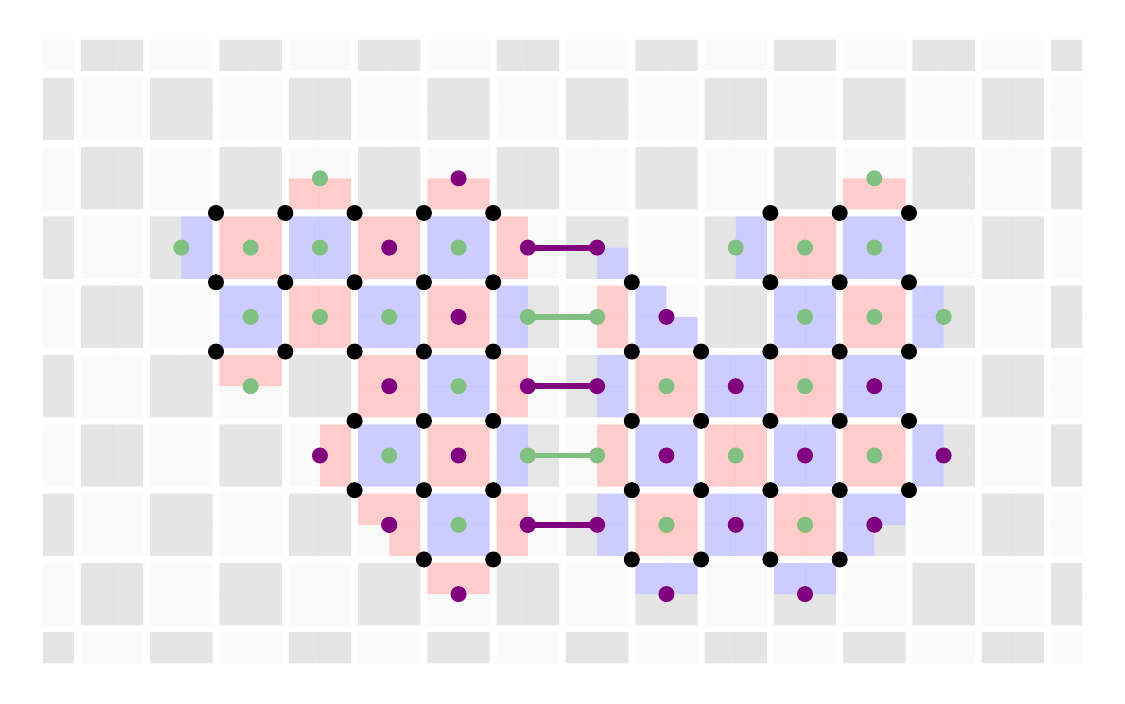}
        \caption{}\label{subfig:need_for_elongated_two_patches_xz}
    \end{subfigure}
    \caption{(a) A QPU with its background lattice coloured grey and white, and two aligned patches laid on top of this lattice. (b) Hence, $XX$ lattice surgery is possible without the need for any long-range stabilisers. The purple auxiliary qubits are the ones whose joint parity outcomes give the logical $XX$-measurement outcome. (c) However, $XZ$ lattice surgery needs some long-range stabilisers, as we need to change the chequerboard pattern to fit the pattern of those stabilisers that multiply into the logical $XZ$. The purple auxiliary qubits' joint parity outcomes gives the logical $XZ$-measurement outcome.}\label{fig:need_for_elongated}
\end{figure}

To define the concept of a background lattice, colour using grey and white a square-grid of the whole QPU via a chequerboard pattern; see \Cref{fig:need_for_elongated}. Now consider a pair of planar surface code patches overlaid on top of this lattice. If both the patches measure $X$ stabilisers on squares of the same colour, we say the patches are aligned (with respect to the background lattice); otherwise, we say they are anti-aligned. For instance, in \Cref{subfig:need_for_elongated_two_patches}, all $X$ stabilisers are laid on top of white squares, hence the two patches are aligned. Between aligned pairs of patches, performing $XX$- or $ZZ$-lattice surgery is possible without any long-range stabiliser measurements. An example of this with $XX$-lattice surgery is depicted in \Cref{subfig:need_for_elongated_two_patches_xx}. The reason for this is that the stabilisers of the merged patch whose product is the logical $XX$ operator (purple dots in \Cref{subfig:need_for_elongated_two_patches_xx}) have to be arranged in a way that forms a chequerboard pattern and, in the case of $XX$ lattice surgery, this is straightforward to achieve. On the other hand, on aligned pairs of patches, performing $XZ$-lattice surgery is not possible without the use of long-range stabilisers, as the required chequerboard pattern for the stabilisers that multiply into logical $XZ$ cannot be made to fit on the chequerboard pattern for stabiliser types we already have on our QPU. However, the use of elongated rectangles makes it possible to measure logical $XZ$, as this introduces a shift in the chequerboard pattern for those stabilisers that multiply into the logical $XZ$; see \Cref{subfig:need_for_elongated_two_patches_xz}. Note that the reverse statements hold for anti-aligned pairs of patches, that is, one can measure $XZ$ and $ZX$ (but not $XX$ or $ZZ$) without elongated rectangles. 

\subsection{Scheduling of tangled rotated planar code patches}\label{app:tangled_middle}
The scheduling we use for the tangled version of rotated planar code patches in \Cref{subsec:rotated_qmem} (\Cref{fig:rotated_patch_sizing}) is convenient, as it can be extended below the bottom boundary in a very natural way. This extension is illustrated in \Cref{subfig:tangled_middle}. More precisely, the scheduling of regular stabilisers below the elongated rectangles is obtained by rotating the scheduling used for regular stabilisers above the elongated rectangles by angle $\pi$. The reader may notice a similar symmetry for component operators as well.

\subsection{Assigning detectors for elongated rectangles and twist defects}\label{app:detectors}
Let us now explain how we assign detectors when performing syndrome extraction for a rotated planar code patch using a set of elongated rectangles and twist defects. Recall that a detector is simply a combination of some measurement outcomes that is deterministic under noiseless execution of the circuit; see \cite{McEwenBaconGidney}. We say that a detector is triggered if this combination of measurements, when sampled from the noisy circuit, is different to the expected deterministic value. Detectors are important because they signal errors in the circuit that we can then correct for.

For planar code patches that contain elongated rectangles and twist defects, we retain the usual detector definitions for local stabilisers. As for the non-local stabilisers, we have three types of detectors, all depicted in \Cref{subfig:tangled_detectors}. One type, $D1$, is the comparison of two non-local stabiliser outcomes from two consecutive rounds where in the latter round we measure the accessory qubits. Hence these detectors consist of four auxiliary qubit outcomes. Type $D2$ detectors are similar, but they compare the non-local stabiliser outcomes from two consecutive rounds where in the former round we measure the accessory qubits. Since the accessory qubits are reset after measurement, their measurement outcomes need to be included as well. As such, $D2$-type detectors consist of six measurement outcomes (four auxiliary qubits and two accessory qubits). The last type of detectors, $D3$, we assign to $Y$-basis accessory qubit measurement outcomes. These outcomes should align with the two Pauli corrections coming from halves of the elongated stabilisers or twist defects, hence they are formed of five measurement outcomes (four auxiliary qubits and one accessory qubit).

\subsection{Effective distance of tangled rotated planar codes}\label{app:rotated-tangled-logical-errors}
In this appendix, we provide an explanation for the effective distance of tangled rotated planar patches.

\begin{figure}
    \centering
    \begin{subfigure}[t]{0.45\textwidth}
        \centering
        \includegraphics[width=\textwidth]{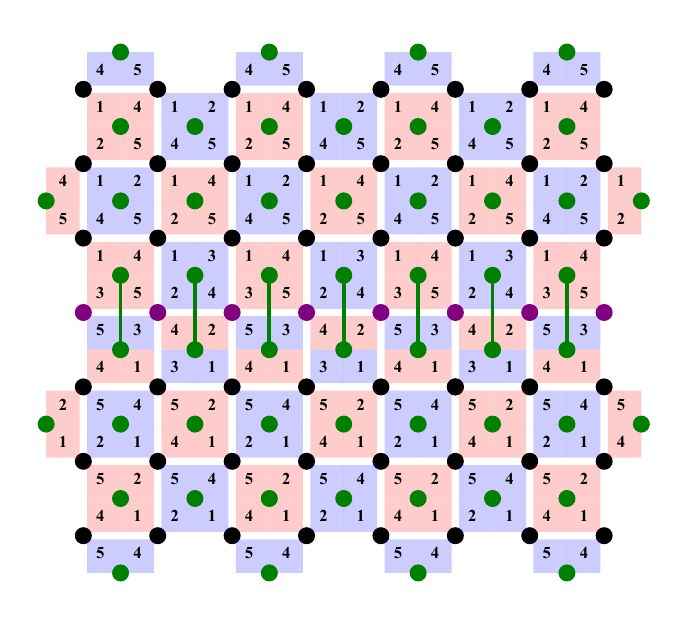}
        \caption{}\label{subfig:tangled_middle}
    \end{subfigure}
    \hfill
    \begin{subfigure}[t]{0.45\textwidth}
        \includegraphics[width=\textwidth]{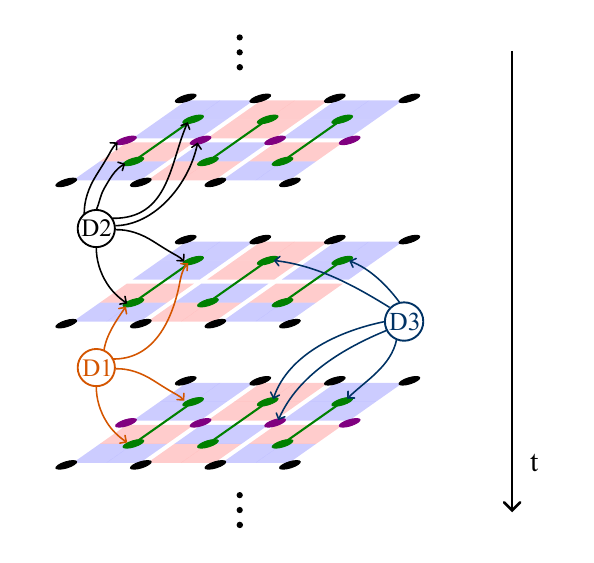}
        \caption{}\label{subfig:tangled_detectors}
    \end{subfigure}
     \caption{(a) The scheduling of the tangled patches in \Cref{fig:rotated_patch_sizing} extends naturally below the elongated rectangles. (b) Three types of detectors that we assign during syndrome extraction with elongated stabilisers and twist defects. $D1-D2$ compares the elongated outcomes from two consecutive rounds, where $D2$ needs to include accessory qubit outcomes, as those qubits were reseted after they were measured. Detector $D3$ compares the $Y$-accessory qubit outcome to the two adjacent Pauli corrections.}\label{fig:tangled_middle}
\end{figure}

We see from the schedules used for the syndrome extraction of regular stabilisers (e.g. in \Cref{subfig:tangled_middle}) that one error on the auxiliary qubit can propagate to an at most weight-$2$ error on data qubits, up to stabiliser equivalence. More precisely, due to the $N$-shaped schedule we use for $X$ stabilisers, this propagated $XX$ error on the data qubits is in the vertical direction, while due to the $Z$-shaped schedule we use for $Z$ stabilisers, the propagated $ZZ$ error is horizontal. Both directions are perpendicular to the direction of the logical operator of the same type. Therefore, each mid-circuit error happening during the syndrome extraction of a regular stabiliser contributes at most weight one towards logical failure.

Next, consider the syndrome extraction of elongated rectangles, and note that similar conclusions apply for twist defects. Note that the qubits in the middle are accessory qubits, hence they are not data qubits of the patch. Therefore, regardless of the schedule, a single mid-circuit error during syndrome extraction for the elongated rectangle can propagate to both data qubits. An example of this is when the Pauli correction $g_1^{m_1+n_1+1}$ is inferred incorrectly from, say, the top component operator due to a classical measurement flip, i.e. we read out $n_1+1$ instead of $n_1$ as the result of a classical measurement readout failure. This means that the Pauli correction is inferred incorrectly as $g_1^{m_1+n_1}$, hence when we apply it in software, we incur an error $g_1$ on the data and accessory qubits. This, however, triggers two types of detectors: 1) the detectors associated with the accessory qubits on which the component operator $g_1$ is supported; and 2) detectors that compare the outcomes of elongated rectangles from two consecutive rounds. Nevertheless, we now have a weight-$2$ data qubit error in the horizontal direction, which is aligned with the $X$ logical operator in the case of $X$-type elongated rectangles.

\begin{figure}
    \centering
    \includegraphics{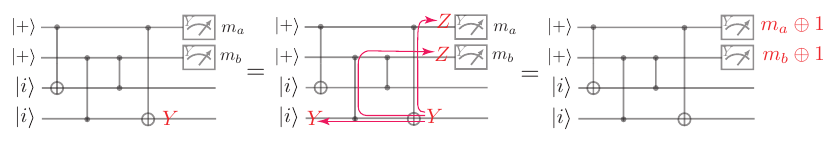}
    \caption{A possible error $Y$ happening mid-circuit:  (left) the original location of the $Y$ error; (middle) propagating to an equivalent Pauli error; (right) the effect of the error on readout.  Since the sum $m_1 \oplus m_2$ is unchanged by the transformations $m_1 \rightarrow m_1 \oplus 1$ and  $m_2 \rightarrow m_2 \oplus 1$, this detector goes untriggered.  However, the Pauli correction after two rounds is inferred from $m_1 \oplus n_1$ and will have the incorrect result. This leads to a Pauli error $g_1$. In the case of the surface code, this leads to two data qubit errors possibly aligned with a logical operator, a so-called bad hook error.  Such a bad hook would also occur if we had directly entangled the auxiliary qubits and so is an intrinsic property of syndrome extraction with low-connectivity hardware. However, this can be compensated by the fact that the distance can be locally increased when we do lattice surgery using the same number of qubits (\Cref{sec:full_lattice_surgery}). Note also that the two detectors associated with the accessory qubits (prepared in the state $\vert i \rangle$) will be triggered after the second round.}
    \label{fig:elongated_hook}
\end{figure}

\begin{figure}
     \centering
     \begin{subfigure}[b]{0.32\textwidth}
         \centering
         \includegraphics[width=1\textwidth]{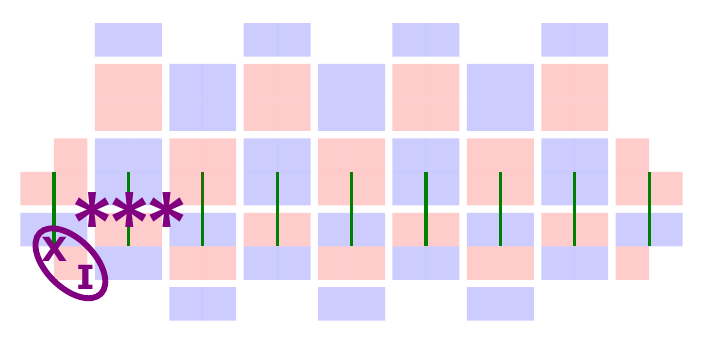}
         \caption{}
         \label{fig:tangled_fault_locations_8x3_1}
     \end{subfigure}
     \hfill
     \begin{subfigure}[b]{0.32\textwidth}
         \centering
         \includegraphics[width=1\textwidth]{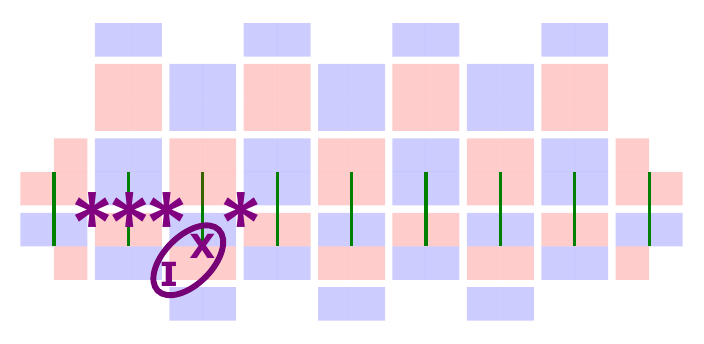}
         \caption{}
         \label{fig:tangled_fault_locations_8x3_2}
     \end{subfigure}
     \hfill
     \begin{subfigure}[b]{0.32\textwidth}
         \centering
         \includegraphics[width=1\textwidth]{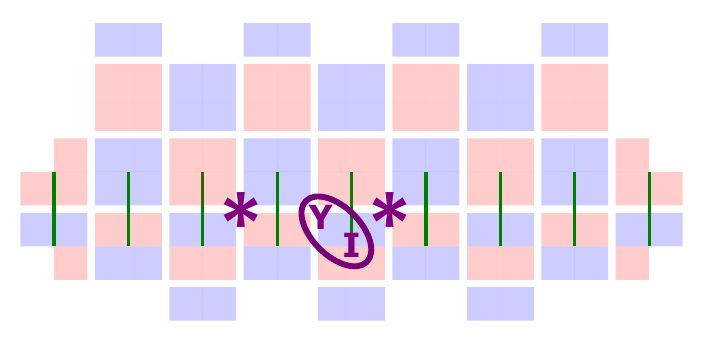}
         \caption{}
         \label{fig:tangled_fault_locations_8x3_3}
     \end{subfigure}
     \centering
     \begin{subfigure}[b]{0.32\textwidth}
         \centering
         \includegraphics[width=1\textwidth]{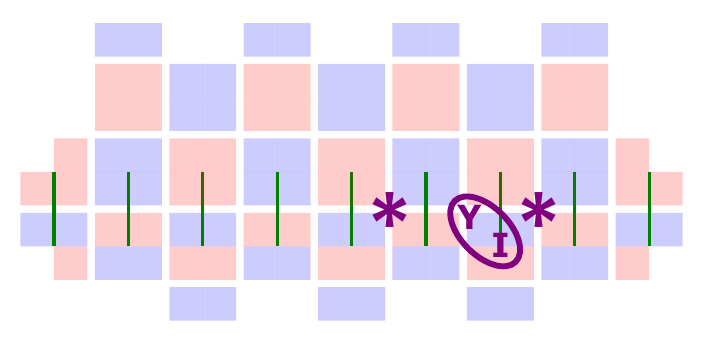}
         \caption{}
         \label{fig:tangled_fault_locations_8x3_4}
     \end{subfigure}
     \hspace{0.02\textwidth}
     \begin{subfigure}[b]{0.32\textwidth}
         \centering
         \includegraphics[width=1\textwidth]{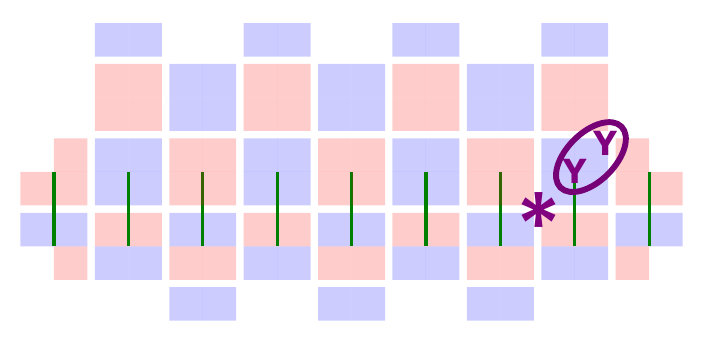}
         \caption{}
         \label{fig:tangled_fault_locations_8x3_5}
     \end{subfigure}
     \hfill
        \caption{Single mid-circuit errors that together lead to a logical $X$ failure. The depicted patch's schedule and qubits are not shown here, but are the same as in \Cref{subfig:tng_patch_8x3}. The mid-circuit errors, shown as purple encircled Pauli strings; each occurs after the execution of a $CZ$ gate. The purple asterisks show which detectors are triggered by the corresponding mid-circuit error. More precisely, if they are placed in the middle of an elongated rectangle, then they are like $D1$--$D2$ from \Cref{subfig:tangled_detectors}; otherwise they are associated with accessory qubits (like $D3$). An exhaustive search shows that fewer than $5$ fault locations do not lead to a logical $X$ failure.}
        \label{fig:tangled_fault_locations}
\end{figure}

Another example of when the Pauli correction is inferred incorrectly is a $Y$-type mid-circuit error happening after the execution of a two-qubit gate; see \Cref{fig:elongated_hook}. The effect of this $Y$ error is that the outcomes of both component operators are flipped. Therefore, again, the Pauli correction is inferred incorrectly. However, this error triggers fewer detectors than the above, namely, the ones associated with the accessory qubit outcomes (cf. \Cref{fig:tangled_fault_locations_8x3_3} or \Cref{fig:tangled_fault_locations_8x3_4}). As above, this also leads to a horizontally-aligned weight-$2$ data qubit error of the same type as the top component operator. 

As the above two types of mid-circuit errors trigger detectors associated with accessory qubits, even in the case where each Pauli correction is inferred incorrectly in either of the above two ways, we do not have an undetectable logical $X$ failure. However, there exists $\left\lfloor\frac{w}{2}\right\rfloor+1$ number of mid-circuit errors that cause an undetectable logical failure; we show an example of this in \Cref{fig:tangled_fault_locations}.

Our observations of the effective distance have been verified using \texttt{stim}~\cite{stim}.

\subsection{General lattice surgery merge-patch with the unrotated planar code}\label{app:unrotated_cut_inside}

\begin{figure}
    \centering
    \includegraphics[width=0.85\textwidth]{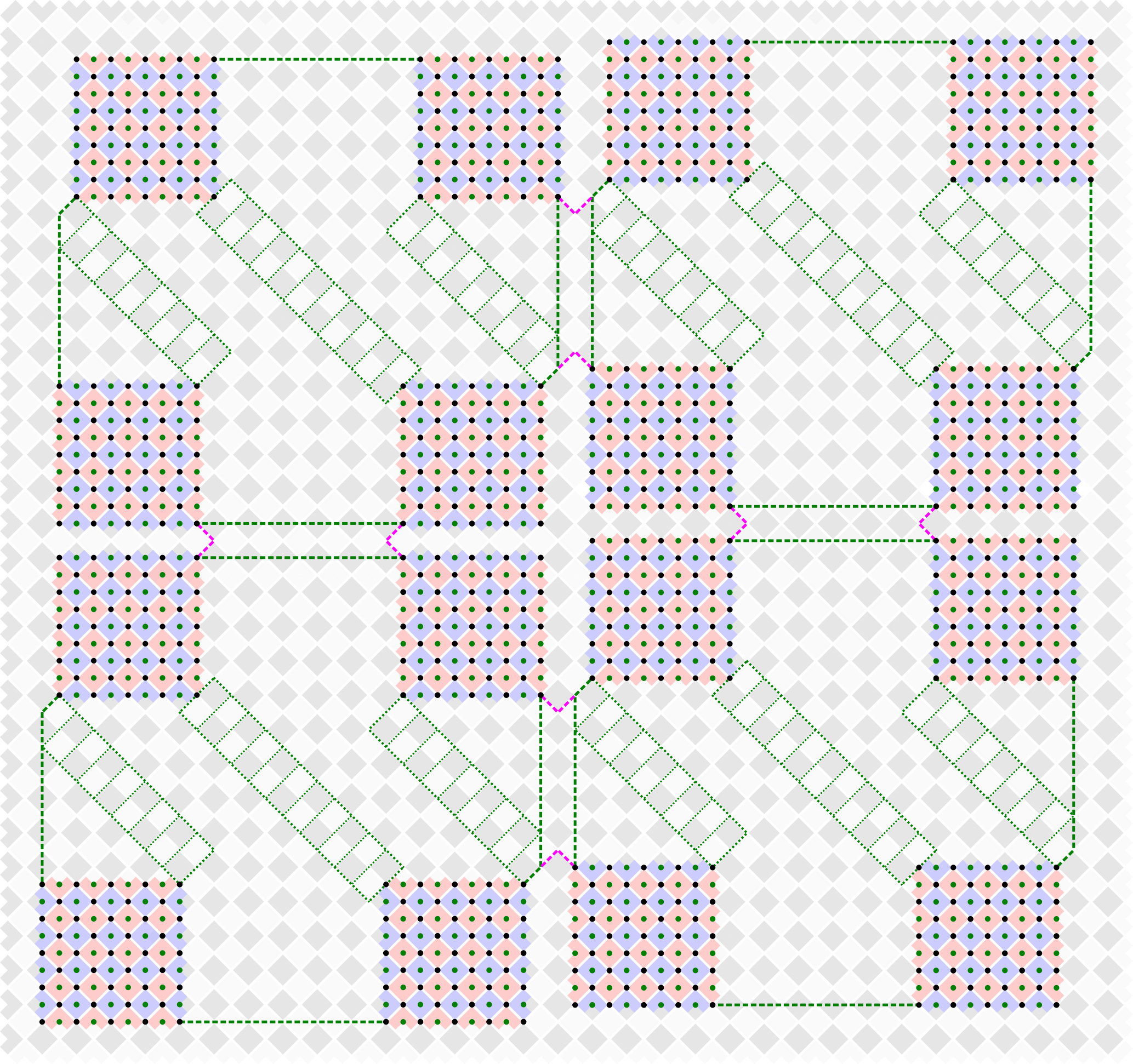}
    \caption{Four tile units in the bulk of the QPU, each surrounded by green dashed lines. The green dotted lines indicate areas where we can place elongated rectangles and twist defects. The purple dashed lines show areas where we may join the merge-stage patches of lattice surgery between tile units. We can continue tiling like this, meaning we shift the patches one row up for the next tile unit to the right.}\label{fig:unrotated_tiling}
\end{figure}

We can tile the QPU in the bulk in a similar way to that in \cite{ChamCamtwistfree}. Namely, we can use tile units each containing qubits that can accommodate $9$ logical qubits, but we only place logical patches at the four corners of each tile unit. \Cref{fig:unrotated_tiling} shows four tile units, each surrounded by green dashed lines. The green dotted lines inside a tile unit show where we may place elongated rectangles and twist defects for the merge stages; otherwise, we place regular stabiliser plaquettes during merge. The purple dashed lines indicate where we may join the merge-stage patches of lattice surgery between tile units.  

\begin{figure}
    \centering
    \begin{subfigure}[b]{0.49\textwidth}
        \centering
        \includegraphics[width=\textwidth]{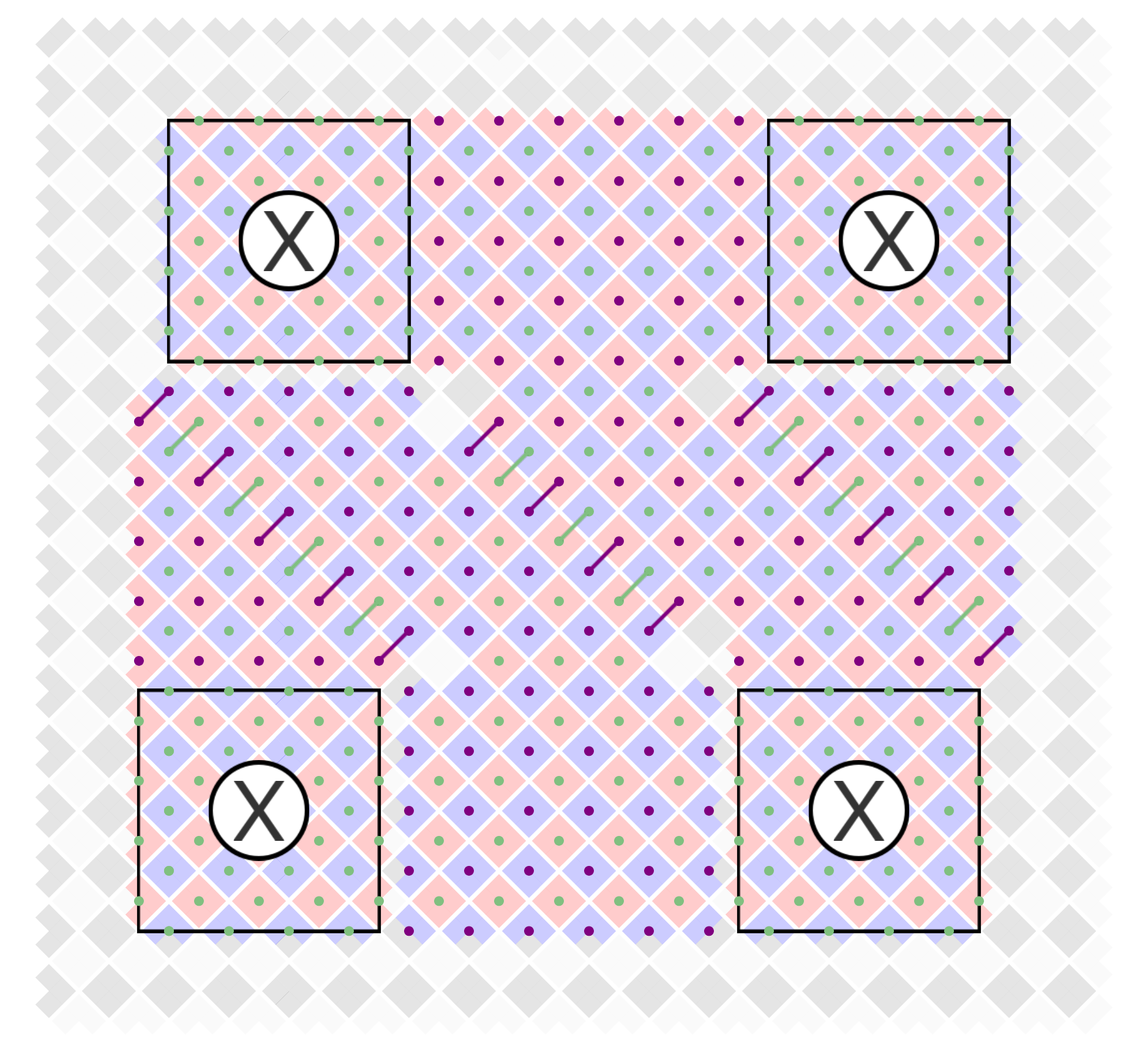}
        \caption{}\label{subfig:unrotated_merge_template_X}
    \end{subfigure}
    \hfill
    \begin{subfigure}[b]{0.49\textwidth}
        \centering
        \includegraphics[width=\textwidth]{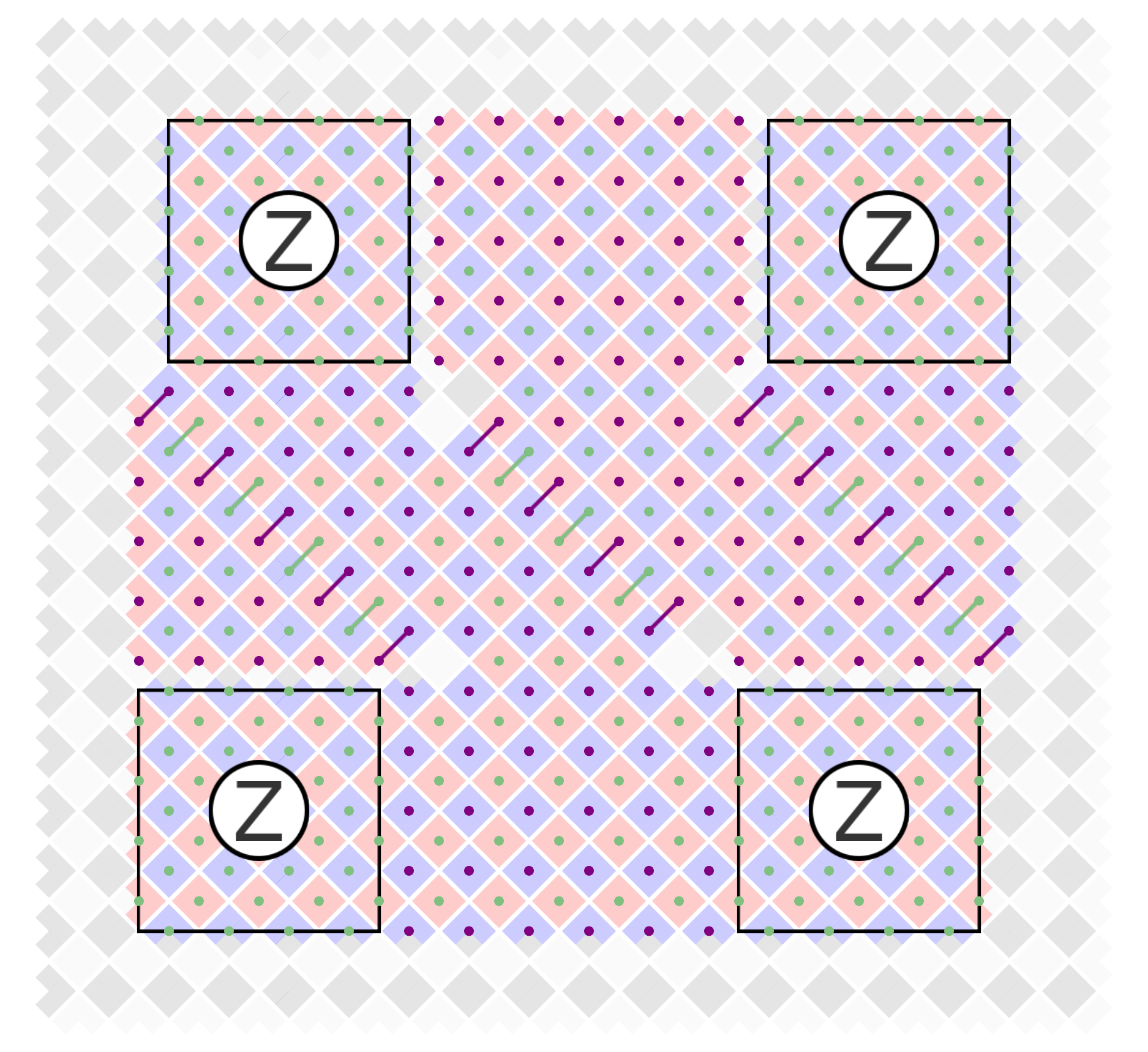}
        \caption{}\label{subfig:unrotated_merge_template_Z}
    \end{subfigure}
    \caption{Lattice surgery examples. (a) Measuring a logical $XXXX$. In general, to measure a logical Pauli product whose term on a patch is $X$, we place plaquettes everywhere except on the grey-white area around that patch in subfigure (a). (b) Measuring a logical $ZZZZ$. In general, to measure a logical Pauli product whose term on a patch is $Z$, we place plaquettes everywhere except on the grey-white area around that patch in subfigure (b).
    If the Pauli term is $Y$, we use all plaquettes around the patch, while if it is $I$, we avoid both grey-white areas. For an explanation of all aspects of the figure, see the caption of \Cref{fig:unrotated_general_lattice_surgery}.}\label{fig:unrotated_merge_template}
\end{figure}

\Cref{fig:unrotated_merge_template} shows a template for constructing a general merge-stage patch inside one tile unit. We start by placing a regular $X$ plaquette on top of each white square, and a $Z$ plaquette on top of each grey square in the tile unit. We then reduce the weight of some of them (possibly removing them completely) according to the template. Namely, if we intend to measure $Y$ on a patch, then we do not do anything around it. In case we wish to measure $X$ on that patch, we reduce the weights of the plaquettes around the patch as shown in \Cref{subfig:unrotated_merge_template_X}. If we would like to measure $Z$ on the patch, then we follow \Cref{subfig:unrotated_merge_template_Z} instead. Finally, if we do not involve the patch in our measurement, then we reduce the weights of the plaquettes around it using both \Cref{subfig:unrotated_merge_template_X} and \Cref{subfig:unrotated_merge_template_Z}. Next, we schedule the remaining plaquettes in a way that two plaquettes are tangled if and only if they are within the same dotted area of \Cref{fig:unrotated_tiling}, so we have pairs of tangled plaquettes. This constructs a general merge-stage patch within one tile unit and, as we argued in \Cref{sec:full_lattice_surgery}, the distance is reduced by at most one, provided we use the standard $N$ and/or $Z$ shaped schedulings everywhere, possibly inserting idling layers to make tangling between some pairs possible.

Now, in order to construct a merge-stage patch that may involve patches from various tile units, we start by performing the construction above within each tile unit. Then, we place weight-$4$ plaquettes in between the purple dashed lines of \Cref{fig:unrotated_tiling} to connect the relevant adjacent tile units. This construction covers the general case, but we note that in many cases there may be room for reducing the number of plaquettes involved, which we do not discuss here, but note in \Cref{sec:full_lattice_surgery}. As an example, we refer to \Cref{fig:unrotated_general_lattice_surgery}, where the construction of the patch was done exactly as described here.

\subsection{A brief discussion on alternatives to FTQC without PBC}
In this paper, we consider the PBC model for FTQC, which executes a quantum algorithm via a series of multi-qubit Pauli measurements. This is a popular model for FTQC, as it enables (virtual) logical Clifford gates with no time cost.  As a consequence, logical $T$ gates are implemented sequentially (they cannot be performed in parallel) and the algorithm runtime is proportional to the $T$-count (more precisely the measurement depth).  This simplicity of algorithm runtime analysis is one of the reasons for the popularity of PBC. Furthermore, assuming only a few magic state factories, PBC is nearly optimal and faster than computational models where the logical Clifford gates are physically performed.  However, in the many-factory limit, the tradeoffs are more subtle and not well understood.

Now, we recall some details on the computational model where the logical Cliffords are physically implemented on a square-grid connectivity QPU. Assume the circuit we are to execute is compiled in terms of Hadamard, $S$, $CNOT$ and $T$ gates. It is well-known that on planar codes the logical $CNOT$ gate can be performed via initialising an auxiliary patch and performing an $XX$ and a $ZZ$ type lattice surgery; see e.g. \cite[Fig. 3]{LitOpp}. Therefore, if the patches are aligned with respect to the background lattice, a logical $CNOT$ gate can be executed on the square-grid connectivity QPU (\Cref{fig:planar_qpu}) using only local stabilisers. If some of these patches are anti-aligned with respect to the background lattice, than we may use tangled schedules to measure some elongated rectangles for these lattice surgery operations. Next, the logical Hadamard gate can be also executed on a square-grid connectivity QPU via patch-deformation; see e.g. \cite{Bombin2021}. It was shown in the same paper (see Fig. 23 there) that the logical $S$ gate can be also performed via patch-deformation on the same QPU. Alternatively, the logical $S$ gate can be performed via initialising an auxiliary patch in a $Y$ eigenstate and performing a $ZZ$-type lattice surgery operation \cite[Fig. 11b]{GoSc} and, recently, improvements to $Y$ state preparation have been made \cite{Gidney2023InplaceAT}. Therefore, the logical $S$ gate can be also performed via a lattice surgery operation on the square-grid connectivity device.  The logical non-Clifford gates are performed using distilled magic states  (similar to PBC), except now the factories can be distributed throughout the architecture.

In summary, we could perform FTQC with the planar code on a square-grid connectivity device, via executing each logical gate sequentially, although the runtime of such approaches is poorly understood compared to the PBC model.

\section{An architecture for fault-tolerant quantum computation with the rotated planar code}\label{app:rotated_lattice_surgery}

In \Cref{sec:full_lattice_surgery}, we presented a way to perform a general multi-qubit logical Pauli measurement with the unrotated planar code on the square-grid connectivity device using one lattice surgery operation. In this section, we present a way to measure a general multi-qubit logical Pauli operator with the rotated planar code that uses one or two lattice surgery operations, depending on the Pauli product we wish to measure. In \cite[Section IV]{ChamCamtwistfree}, the authors describe a scheme, called twist-free lattice surgery, that uses two lattice surgery operations for each multi-qubit Pauli measurement that involves a $Y$ term, thereby roughly doubling the time-cost. For that scheme to work, the authors introduced a so-called dislocation area in the square-grid connectivity hardware (see \cite[Fig. 11]{ChamCamtwistfree}), whose role is effectively to switch the background lattice. The dislocation area imposes constraints on the merge-stage patch of the lattice surgery operations; in particular, they need to include an auxiliary patch that is initialised close to the dislocation area, and the second merge-stage patch needs to include the dislocation area itself. This may result in needing large merge-stage patches even in the case when, e.g. we measure a Pauli product on patches that are close to each other but far away from the dislocation area.

In this section, we present an alternative protocol that does not need the dislocation area, hence can be performed on the square-grid connectivity hardware, and we will argue that the space-time-cost remains approximately the same as the twist-free protocol of \cite{ChamCamtwistfree}. We achieve this by noticing that certain multi-qubit Pauli measurements need only one twist-based lattice surgery operation that can be performed with tangled schedules, and we perform the rest of the Pauli measurements by mimicking the the twist-free protocol of \cite{ChamCamtwistfree} but with using tangled schedules instead of the dislocation area.

\begin{figure}
    \centering
    \includegraphics[width=0.85\textwidth]{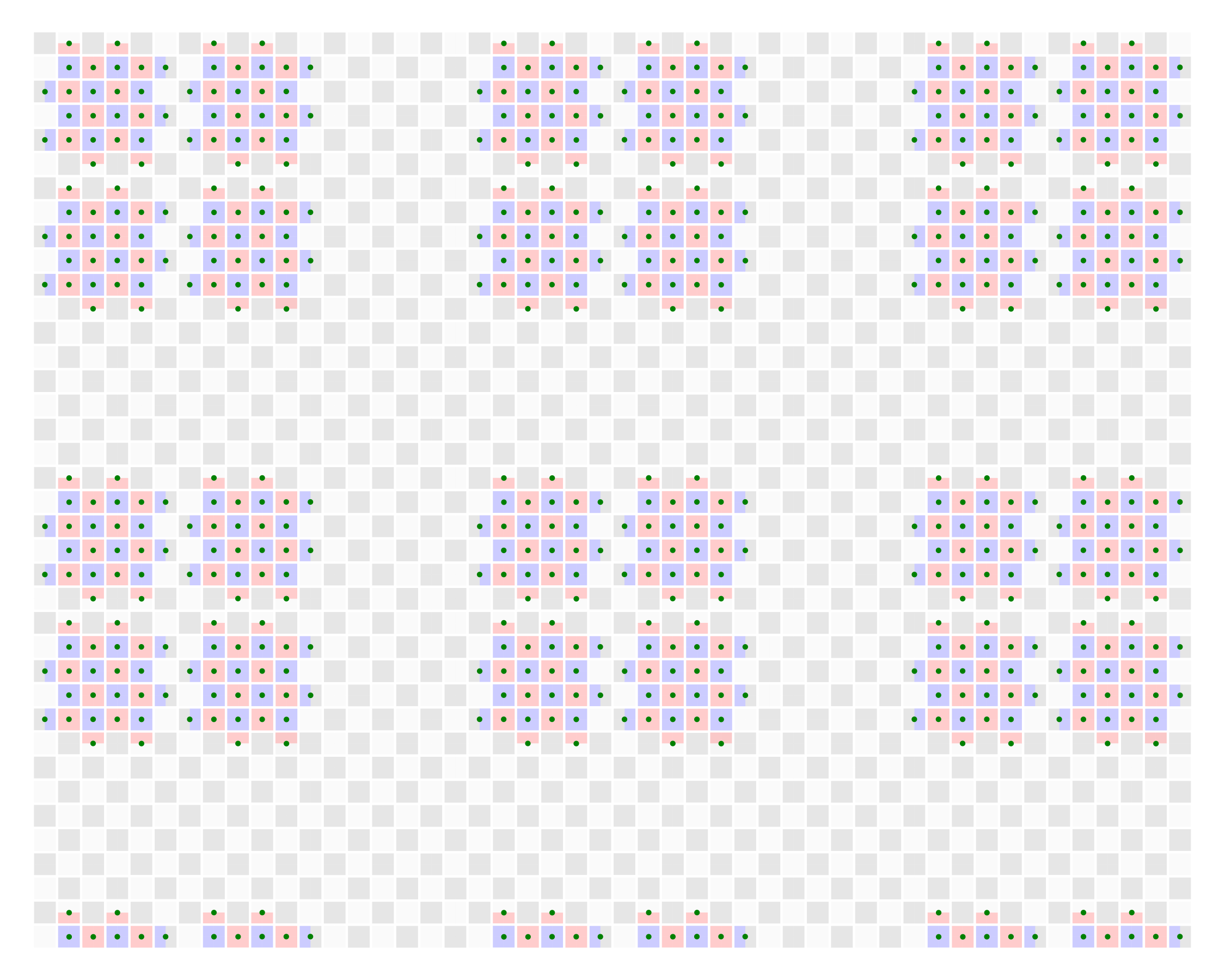}
    \caption{The tiling of our QPU in the bulk with rotated planar code patches.}
    \label{subfig:rotated_qpu_tiling}
\end{figure}

We place the logical patches in a similar way to \cite[Fig. 11]{ChamCamtwistfree}, which is illustrated in \Cref{subfig:rotated_qpu_tiling}. There are two differences, one being that, for sake of simplicity, we consider square-shaped patches with equal $X$ and $Z$ distances. The other difference is that we make the horizontal and vertical parts of the routing space two data qubits taller and wider, respectively. Thus we use an additional $O(nd)$, qubits where $n$ denotes the number of logical patches, which is negligible compared to the already needed $O(nd^2)$ qubits. Therefore, we can consider the space-cost to be approximately the same. Padding the routing space this way, however, makes it possible to connect the logical patches during merge stages via both the horizontal and vertical parts of the routing space, which is not the case in \cite[Fig. 11 c)-d)]{ChamCamtwistfree}. Therefore, merge-stage patches are potentially able to be smaller.

\begin{figure}
      \centering
      \begin{subfigure}[b]{\textwidth}
          \centering
          \includegraphics[width=0.69\textwidth]{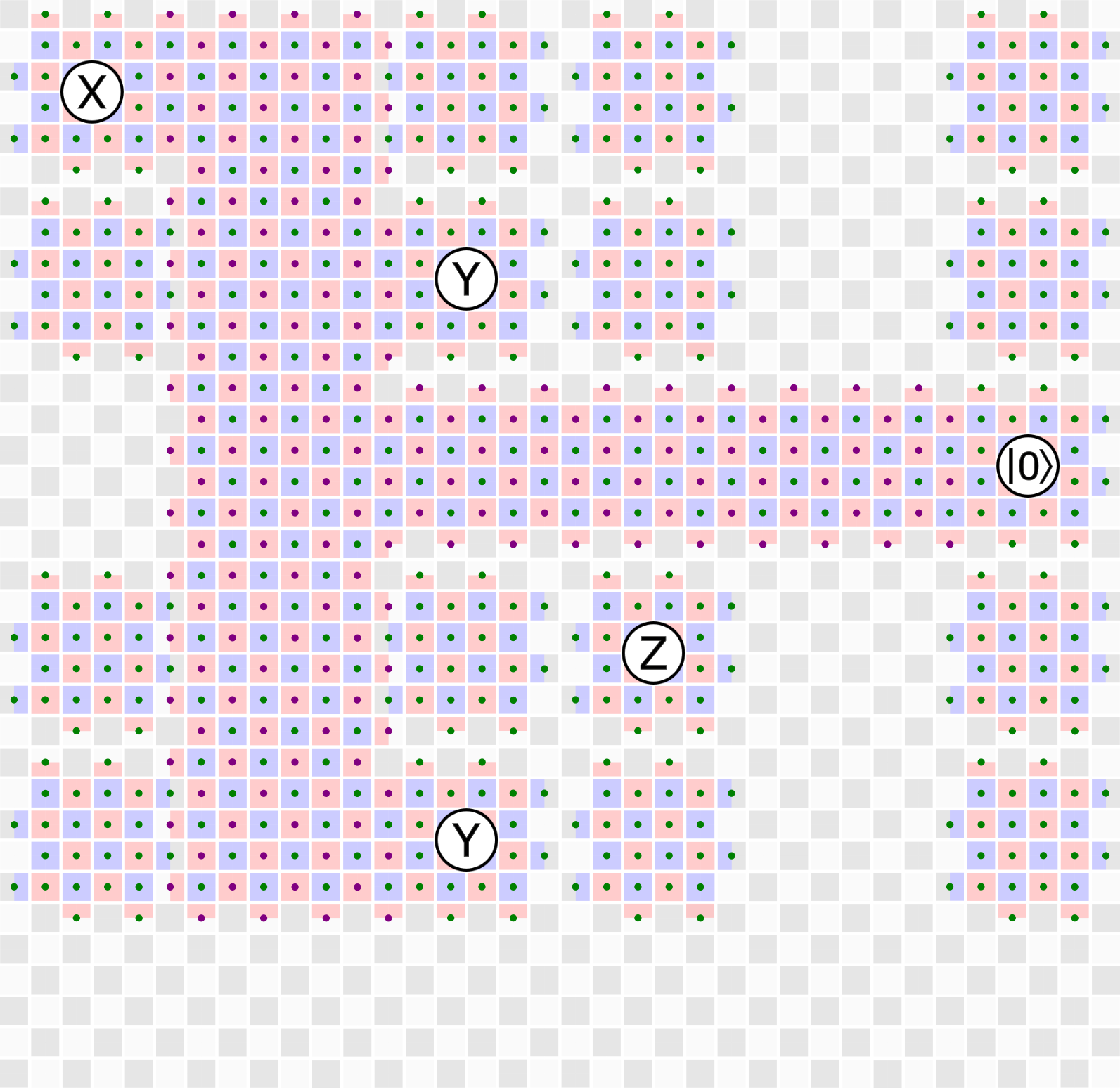}
          \caption{The first, purely $X$-type, lattice surgery operation involving the three patches on which we intend to measure either $X$ or $Y$, and the auxiliary patch. Local stabilisers are sufficient here.}
          \label{subfig:rotated_twist_free_3}
      \end{subfigure}
      \vfill
      \centering
      \begin{subfigure}[b]{\textwidth}
          \centering
          \includegraphics[width=0.69\textwidth]{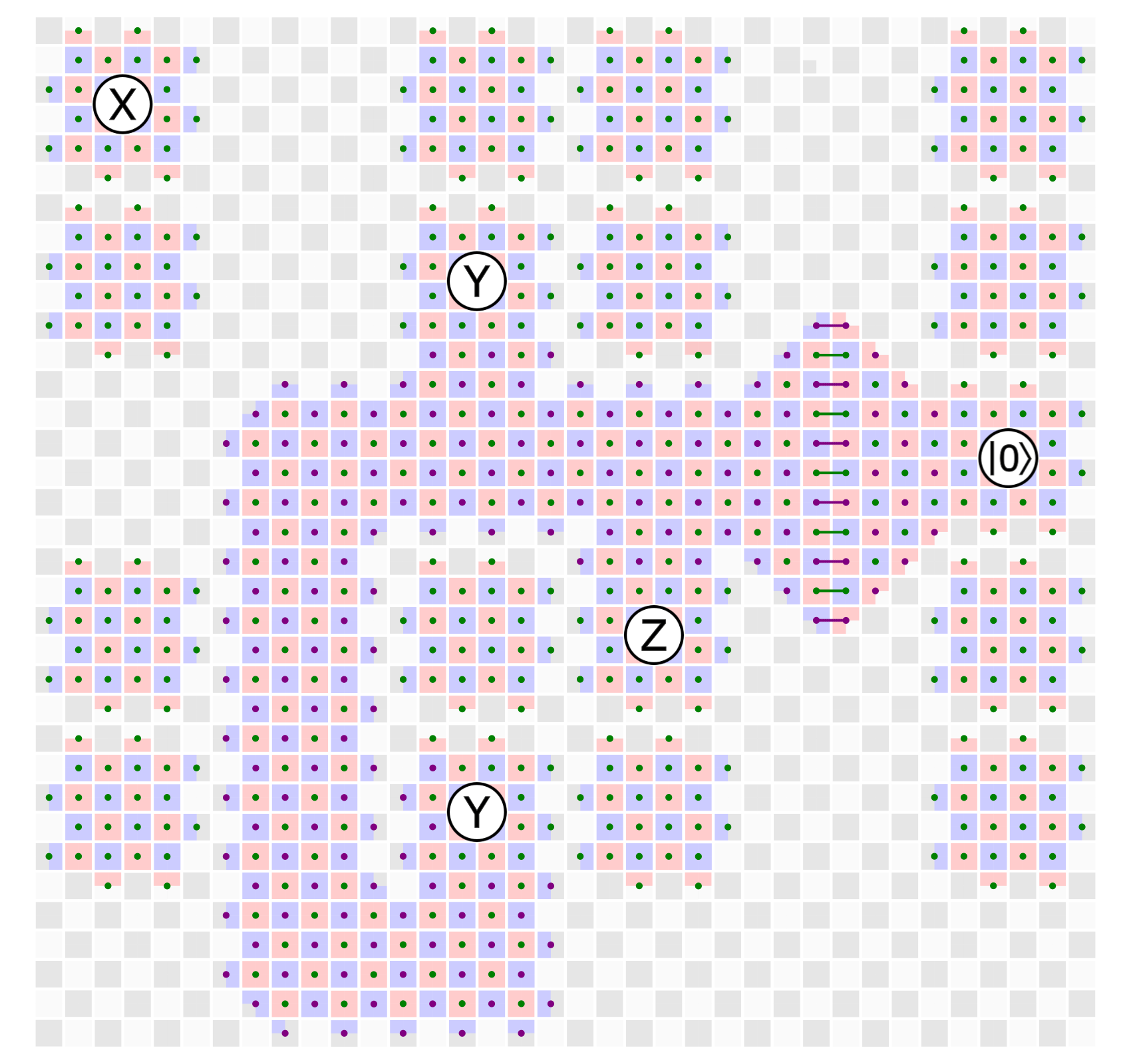}
          \caption{The second lattice surgery operation involving the three patches on which we intend to measure either $Y$ or $Z$, and the auxiliary patch. This is a mixed $ZZZX$-type lattice surgery, hence we need elongated rectangles. Due to the distance halving effect, we need to have a wider area to connect the auxiliary patch to the rest of the patches.}
          \label{subfig:rotated_twist_free_4}
      \end{subfigure}
         \caption{Performing the same $XYZY$ measurement as in \cite[Fig. 11]{ChamCamtwistfree} but on the degree-four hardware \Cref{fig:planar_qpu} using rotated planar patches.}
         \label{fig:rotated_twist_free}
 \end{figure}

Now, as an illustrative example, we perform the same $XYZY$-measurement from \cite[Fig. 11]{ChamCamtwistfree} with our tangled schedules method, first via twist-free, and then via twist-based lattice surgery. In \Cref{subfig:rotated_twist_free_3}, a $|0\rangle$-state auxiliary patch is initialised in the bulk-routing space, between two logical patches, and we perform an $XXIXX$-type lattice surgery, where we only need to use regular local range stabilisers. This is then followed by performing the $IZZZX$-type lattice surgery shown in \Cref{subfig:rotated_twist_free_4}, for which we need to use elongated rectangles. In particular, due to the distance-halving effect of the tangled schedules method (when using the rotated planar code), we need to widen a part of the merge-stage patch. Finally, we measure out the auxiliary patch in the $Z$ basis and, depending on the outcome, we may apply a Pauli correction in software; the full details are in \cite[Section IV]{ChamCamtwistfree}. Through this example, the reader can see that indeed this twist-free protocol works for measuring a general Pauli product. 

We note that there is an alternative to the second lattice surgery step in the twist-free case, that is not a possibility in the rest of the cases considered in the paper. Namely, we could swap the auxiliary patch to the left with one column of data qubits using two layers of $SWAP$ gates (which compile to $6$ layers of $CZ$ gates and some further layers of one-qubit gates, if $SWAP$ is not a native gate), and then as we now have the auxiliary patch on a different background lattice, it is enough to use regular stabilsers for the merge step of the second lattice surgery operation. This solution could be applied also in the case of \cite{ChamCamtwistfree} instead of introducing a dislocation area. Even though this adds further noise to the lattice surgery operation, in some noise regimes it may be beneficial to use this method. However, in general we expect this alternative not to be the best option.

\begin{figure}
      \centering
          \includegraphics[width=1\textwidth]{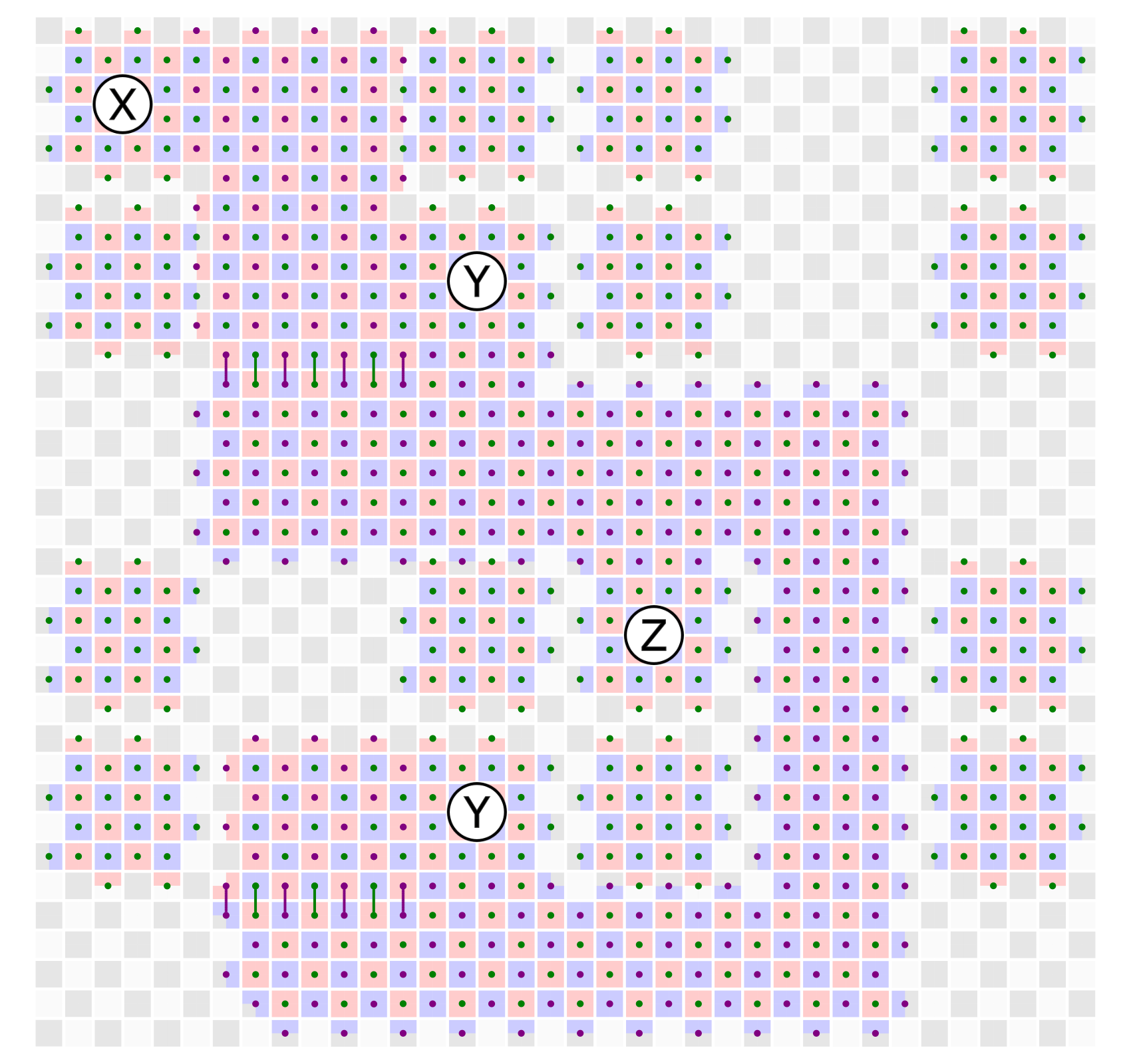}
          \caption{An alternative way to measure the $XYZY$ operator of \Cref{fig:rotated_twist_free} with twist-based lattice surgery with rotated planar patches. While twist-based lattice surgery is not always possible in this way without losing significant distance, in several cases it is possible.}
         \label{fig:rotated_twist_based}
 \end{figure}

An advantage of our tangled syndrome extraction technique is that often we do not need to perform two lattice surgery operations to measure the Pauli product. Indeed in many cases it turns out that we can actually do it with just one lattice surgery operation. This is the case, for instance, for the $XYZY$-measurement of \cite[Fig. 11]{ChamCamtwistfree}; we illustrate how to do this in \Cref{fig:rotated_twist_based}.

Next, we estimate the time cost of the mixed method, namely, when we use twist-free lattice surgery whenever the twist-based version is not possible. Assume that we need $N$ Pauli measurements and $qN$ of these can be done with the twist-based version, where $0\leq q\leq 1$. Then, our time-cost is $\frac{qR + (1-q)(1+R)}{2}$-times the time-cost of twist-free lattice surgery from \cite{ChamCamtwistfree}, where $R$ is the time-overhead discussed in \Cref{subsec:stability}. For instance, in case $q=\frac{1}{2}$, this multiplicative factor is less than $1$ if and only if the time overhead $R$ is less than $1.5$. Recall from \Cref{subsec:stability} that we expect this would be the case for larger patches that typically arise here, and further we would expect the time overhead to decrease with the improvement of physical error rates. Therefore, in general we expect to achieve a similar space-time cost for fault-tolerant computation with tangled schedules to that with the twist-free protocol of \cite{ChamCamtwistfree}. Furthermore, we also removed the need to modify the hardware.

Finally, we compare the space cost (i.e. qubit count) of our scheme to some previous schemes from the literature. We already mentioned that our scheme and the one from \cite{ChamCamtwistfree} have approximately the same space cost, i.e. $n$ logical qubits can be stored with $\approx\frac{9}{4}n \cdot (2d^2+O(d))$ qubits, where $d$ is the distance of logical patches. We further compare our scheme to the three schemes presented in \cite{GoSc}: ``compact block'', ``intermediate block'' and ``fast block''. As hardware connectivity constraints were not taken into account in \cite{GoSc}, we assume for this comparison that for each stabiliser we always have an auxiliary qubit available which is connected to its data qubits. Then the compact block needs $(\frac{3}{2}n + o(n)) \cdot (2d^2+O(d))$, while the intermediate block and fast block both need $(2n+o(n)) \cdot (2d^2+O(d))$ qubits to store $n$ logical qubits. Our scheme, which takes into account the hardware connectivity constraints, requires approximately $1.5\times$ the space cost of the compact block and $1.125\times$ the space cost of the intermediate and fast blocks.

\section{Proof of \Cref{thm:main}}\label{app:proof_of_main_thm}

\begin{figure}
\includegraphics[width=0.95\columnwidth]{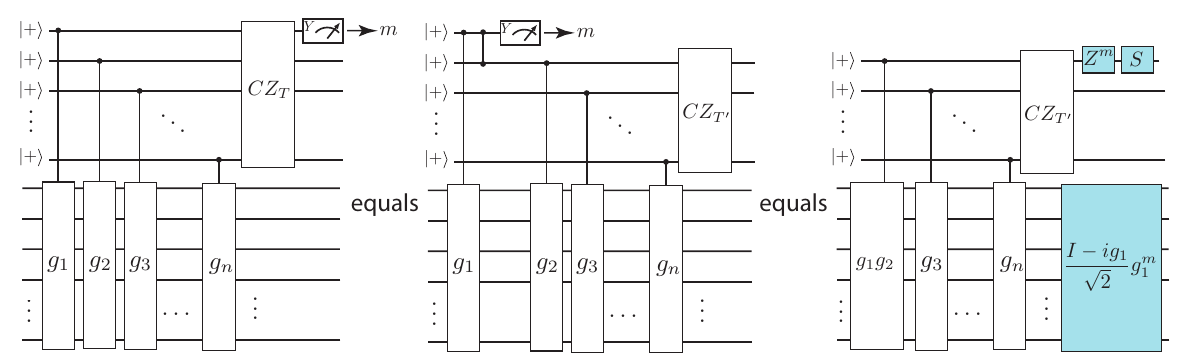}
\caption{Pruning a single leaf from a tree of tangled schedules. As a result of pruning, on the right, we see a Clifford correction on data qubits and additional one-qubit gates on auxiliary qubit $2$ (gates in blue boxes).} \label{fig:Leaf_from_tree}
\end{figure}

Here, we prove \Cref{thm:main}. For each pair of schedules that are tangled, we can reorder the gates (as we did in \Cref{fig:2_plaqs_circ}) so that all gates involving a particular auxiliary qubit are adjacent. If component circuit $\mathcal{C}_j$ is tangled with component circuit $\mathcal{C}_k$, we gain a term $CZ_{j,k}$ between the relevant auxiliary qubits. Therefore, the reordered circuit includes a block of $CZ$ gates; see the leftmost circuit in \Cref{fig:Leaf_from_tree}. This $CZ$-block is exactly
\begin{equation}
  CZ_G = \prod_{(j,k)\in E} CZ_{j,k},
\end{equation}
where the graph $G = (V,E)$ is defined in the statement of the theorem in the main text. Our theorem assumes that $G$ is a forest, i.e. a disjoint union of trees (where a tree is a connected graph with no cycles).

First, we prove our theorem for the case when $G$ is a single tree, and write $T$ for the graph in this case to emphasise this. For convenience, we label the vertices with integers from 1 to $m$, where $m$ is the number of vertices in the tree. Recall that we wish to measure the operator $h = \prod_{j=1}^{m}g_{j}$, where each $g_{j}$ is a multi-qubit Pauli component operator and $[g_{j}, g_{k}]=0$ holds for all $j, k$.

We identify a leaf vertex, i.e. one that is connected to only one other vertex by an edge. We may assume that this leaf is vertex $1$ and it is connected (via an edge) to vertex $2$. Then, we can decompose the $CZ$-block as
\begin{equation}
  CZ_{T} = CZ_{T'} CZ_{1,2},
\end{equation}
where $T'$ is the tree obtained by removing the vertex $1$ and the edge $(1,2)$. Now, via the pruning identity (\Cref{Fig:Prune}), it is straightforward to see the equivalences of the three circuits depicted in \Cref{fig:Leaf_from_tree}. Note that, as the correction on the data qubits is a function of $g_1$, it commutes with all other component operators and thus can be moved to the right as shown in the figure. 

From here, we will iteratively identify a leaf of the current tree and will use the pruning identity to prune that leaf from the tree. Assume that, after some sequence of prunes, we have a smaller tree, $\tilde{T}=(\tilde{V},\tilde{E})$.  Let $v \in \tilde{V}$ denote the next leaf vertex in $\tilde{T}$ to be pruned.  Let $\mathcal{N}_{T, \tilde{T}}(v)$ denote the set of vertices in the original tree $T$ such that $u \in \mathcal{N}_{T, \tilde{T}}(v)$ if and only if
\begin{enumerate}
    \item $u$ was originally a neighbour of $v$; formally, $(u,v)\in E$; and
    \item $u$ has since been pruned; formally, $u \notin \tilde{V}$.
\end{enumerate}
Then, it follows that the accumulated one-qubit gates on vertex $v$ have come from applications of the pruning identity on each $u \in \mathcal{N}_{T, \tilde{T}}(v)$. Therefore, the form of the total accumulated one-qubit gates is
\begin{equation}
    W(v) = S^{|\mathcal{N}_{T, \tilde{T}}(v)|} Z^{\sum_{u \in \mathcal{N}_{T, \tilde{T}}(v)} m_u},
\end{equation}
 where $m_{u}$ is the outcome of measuring qubit $u$. Note that, since $v$ is a leaf in $\tilde{T}$, it only has a single remaining neighbouring vertex. Letting $d(v)$ denote the vertex degree of $v$ with respect to the original tree $T$, we have that $d(v) = 1 + |\mathcal{N}_{T, \tilde{T}}(v) |$.  Introducing the shorthand $M(v):=\sum_{u \in \mathcal{N}_{T, \tilde{T}}(v)} m_u$, we get the statement that
\begin{equation}
    W(v) = S^{d(v)-1} Z^{M(v)}.
\end{equation}
So that we can apply the pruning identity again, we measure qubit $v$ so that we cancel the accumulated one-qubit gates and then measure in the Y basis. Therefore, we measure qubit $v$ in the conjugated basis
\begin{equation}
    S^{d(v)-1} Z^{M(v)} Y Z^{M(v)} S^{1 - d(v)} = 
    \begin{cases}
        (-1)^{M(v)+\frac{d(v)-1}{2}} \, Y & \mathrm{if} \ d(v) \ \mathrm{odd}, \\
        (-1)^{M(v)+\frac{d(v)}{2}} \, X & \mathrm{if} \ d(v) \ \mathrm{ even}.
    \end{cases}
\end{equation}
Note that the minus signs can be handled in classical post-processing.

Now, if we apply the pruning lemma with leaf $v$, we obtain an additional one-qubit gate on the auxiliary qubit that was connected to $v$ in $\tilde{T}$. Further to that, the pruning also leads to a Clifford correction on the data qubits. It is straightforward to see that, after using the pruning identity for the $j^{\mathrm{th}}$ time, the total incurred Clifford correction on the data qubits has the form 
\begin{equation}
    \prod_{k=1}^{j} V_k P_k^{m_k},
\end{equation}
where $V_k = \frac{1}{\sqrt{2}}(I-iP_k)$, $m_k$ is the outcome corresponding to the auxiliary qubit that was pruned during the $k^{\mathrm{th}}$ step, and $P_k$ is a product of some component operators. (For instance, if $2$ is a leaf of $T'$ after the first pruning step, we may identify $2$ as the next leaf, and then $P_2=g_1g_2$.) Recall that the component operators commute with each other, hence the full correction can be moved to the right of the circuit.

Let us assume we have done $m-1$ pruning steps, so that only one vertex, say $w$, remains. Then, the state on the data qubits and qubit $w$ has the form
\begin{equation}
   \ket{\Psi}  \propto W(w) \cdot C_w (h) \cdot \left[ \ket{+} \otimes \prod_{k=1}^{m-1} V_k P_k^{m_k} \ket{\psi}  \right],
\end{equation}
where $C_w(h)$ denotes the controlled Pauli $h$ gate, with control being on $w$, $h = \prod_{j=1}^{m}g_{j}$ and $\ket{\psi}$ is the initial state on the data qubits. We now measure $w$ in the $W(w) \, X \, W(w)^\dagger$ basis though, again, it suffices only to measure in the correct basis up to sign, as this sign may be applied classically. We see that performing such a measurement and then applying the correction $\left(\prod_{k=1}^{m-1} V_k P_k^{m_k}\right)^\dagger$ to the data qubits is equivalent to measuring the Pauli operator $h$ with outcome $m_w$.

We note that, up to sign, the bases of all auxiliary qubit measurements can be determined ahead of time as these depend only on the degrees of the corresponding vertices. The signs can be applied classically when determining the outcome of measuring the operator $h$ and the Clifford correction. Therefore, all auxiliary qubits can be measured simultaneously.

We conclude the proof by noting that, in case $G$ is a forest, we can prune each tree as above. Each correction on data qubits can be moved to the right, as the component operators commute with each other. Finally, we end up with only having single vertices in our graph, where the measurements of each remaining qubit will be equivalent to measuring the product of those component operators which were in the tree of that measured qubit in the original graph $G$.

\end{document}